\documentclass[11pt,a4paper,aps,preprint,superscriptaddress,nofootinbib]{revtex4-1}
\usepackage{graphicx}
\usepackage{amssymb}
\usepackage{amsmath}
\usepackage{multirow}
\usepackage{color}
\usepackage[colorlinks=true, pdfstartview=FitV, linkcolor=blue, citecolor=blue, urlcolor=blue]{hyperref}
\usepackage[utf8]{inputenc}

\newcommand{\ket}[1]{\ensuremath{\left|#1\right\rangle}}
\newcommand{\bra}[1]{\ensuremath{\left\langle#1\right|}}
\newcommand{\braket}[1]{\ensuremath{\left\langle#1\right\rangle}}
\begin{document}

\title{Sensitivity of future lepton colliders and low-energy experiments to charged lepton flavor violation from bileptons}

\author{Tong Li}
\email{litong@nankai.edu.cn}
\affiliation{
School of Physics, Nankai University, Tianjin 300071, China
}
\author{Michael A.~Schmidt}
\email{m.schmidt@unsw.edu.au}
\affiliation{
School of Physics, The University of New South Wales, Sydney, New South Wales 2052, Australia
}

\begin{abstract}
The observation of charged lepton flavor violation is a clear sign of physics
beyond the Standard Model (SM).
In this work, we investigate the sensitivity of
future lepton colliders to charged lepton flavor violation via on-shell
production of bileptons, and compare their sensitivity with current constraints and future sensitivities of low-energy experiments.
Bileptons couple to two charged leptons with possibly different flavors and are obtained by expanding the general SM gauge invariant Lagrangians with or without lepton number conservation.
We find that future lepton colliders will provide complementary sensitivity to the charged-lepton-flavor-violating couplings of
bileptons compared with low-energy experiments. The future improvements of
muonium-antimuonium conversion, lepton flavor non-universality in leptonic $\tau$ decays, electroweak precision observables and the anomalous
magnetic moments of charged leptons will also be able to probe similar
parameter space.
\end{abstract}

\maketitle

\section{Introduction}
\label{sec:Intro}

The observation of neutrino oscillations and thus non-zero neutrino masses clearly
established the existence of lepton flavor violation in the neutrino sector. We also expect the existence of charged lepton flavor violation (CLFV) which occurs in short-distance processes without neutrinos in the initial or final state. In the Standard Model (SM) with three massive neutrinos, the rates
of CLFV processes are suppressed by $G_F^2 m_\nu^4\lesssim10^{-50}$ due to unitarity of the leptonic mixing matrix~\cite{Cheng:1980tp} and thus beyond the sensitivity of any
current or planned experiments. Hence, the observation of any CLFV process
implies the existence of new physics beyond the SM with three massive
neutrinos.
The CLFV is predicted by many different new physics models (see
Refs.~\cite{Lindner:2016bgg,Calibbi:2017uvl} for recent reviews), including neutrino mass models
such as the inverse seesaw model~\cite{Tommasini:1995ii} and radiative neutrino
mass models~\cite{Cai:2017jrq}. It may also arise in other extensions of the SM
such as the multi-Higgs doublet models~\cite{Branco:2011iw} or the minimal
supersymmetric SM via gaugino-slepton loops with off-diagonal terms in the
slepton soft mass matrix~\cite{Raidal:2008jk,Calibbi:2017uvl}.

As CLFV induces rare processes, they are generally searched at
low-energy experiments with high intensity. See
Refs.~\cite{Carpentier:2010ue,Petrov:2013vka} for a list of constraints on
the effective CLFV operators obtained from several low-energy precision
measurements. The CLFV processes may also be searched for at high-energy colliders.
The Large Electron-Positron Collider (LEP) sets upper limits on the branching ratio of $Z$ boson rare decays~\cite{Tanabashi:2018oca}, i.e. $Z\to \ell \ell'$ induced by loop
diagrams, and still provides the most stringent constraint on the branching
ratios of $Z\to \tau e(\mu)$ as $0.98 (1.2)\times 10^{-5}$ up to now. The ATLAS experiment currently
sets the most stringent limit of $7.3\times 10^{-7}$ on BR($Z\to
e\mu$)~\cite{Aad:2014bca} and comparable limit of $5.8 (2.4)\times 10^{-5}$ on
BR($Z\to \tau e(\mu)$)~\cite{Aaboud:2018cxn}. The future $Z$ factories could
improve the sensitivity by about four orders of
magnitude~\cite{CEPCStudyGroup:2018ghi}. The CLFV can also occur in Higgs
boson decay through the dimension-6 operator $H^\dagger H \bar L e_R H$ in SM
effective field theory~\cite{Harnik:2012pb}. The Large Hadron Collider (LHC) recently improved its limit on
the effective $\tau\ell$ couplings to the level of $(1-2)\times
10^{-3}$~\cite{Sirunyan:2017xzt,ATLAS:2019icc} and the proposed Higgs boson
factories are expected to be sensitive to CLFV couplings down to the
order of $10^{-4}$ ~\cite{CEPCStudyGroup:2018ghi,Qin:2017aju}.

Besides these rare decays, CLFV can also be probed through scattering processes
at colliders. A hadron collider is sensitive to effective operators with two
colored particles and two leptons in processes such as $q\bar q\to \ell
\ell^\prime$~\cite{Cai:2015poa} and $gg\to
\ell\ell^\prime$~\cite{Bhattacharya:2018ryy,Cai:2018cog}. The hadron colliders are also sensitive to a number of higher dimension operators contributing to anomalous triple or quartic gauge couplings through $WW$ scattering process $pp\to 2 \ {\rm jets}+W^\ast W^\ast\to 2 \ {\rm jets}+\ell \nu_\ell \ell' \nu_{\ell'}$~\cite{Kalinowski:2018oxd,Chaudhary:2019aim}. A lepton collider may
probe effective operators with four charged leptons via $e^+ e^- \to \ell \ell^\prime$.
These searches can also be interpreted in terms of simplified models. CLFV processes at the lepton collider can be described by seven bileptons~\cite{Cuypers:1996ia}, which are scalar or vector bosons coupled to two leptons via a renormalizable coupling. In particular, off-shell bileptons can mediate the processes $e^+ e^- \to \ell \ell^\prime$ whose potential observation at future lepton colliders has recently been studied by us~\cite{Li:2018cod}. See also Ref.~\cite{Dev:2017ftk} and Refs.~\cite{Sui:2017qra,Dev:2018upe} for related studies of electroweak
doublet and triplet scalar bileptons, respectively.
Another promising probe for CLFV is through the on-shell production of a bilepton $X$ together with two charged leptons with different flavors, i.e. $e^+ e^- \to X \ell \ell^\prime$. This production scenario only depends on a single CLFV coupling in each production channel and thus can be directly compared with other constraints.
On-shell production has been studied in Ref.~\cite{Dev:2017ftk} for an electroweak doublet scalar and Refs.~\cite{Sui:2017qra,Dev:2018upe} for an electroweak triplet scalar.

The main aim of this work is to explore the sensitivity reach to the CLFV couplings for all seven bileptons through the on-shell production of a bilepton $X$ in association with two charged leptons at proposed future lepton colliders. We compare the sensitivities of future lepton colliders with the existing constraints and future sensitivities of other experiments. Currently, the most relevant constraints are from the anomalous magnetic moments (AMMs) of electrons and muons, muonium-antimuonium conversion, lepton flavor universality (LFU) in leptonic $\tau$ decays, electroweak precision observables, and previous collider searches at the LEP and the LHC experiments.
Our analysis here goes beyond the previous work by extending the study to all possible bileptons and including additional constraint from the violation of lepton flavor universality in leptonic $\tau$ decays and electroweak precision observables as well as a discussion of neutrino trident productions. We also improve the calculation of muonium-antimuonium conversion for the bileptons.

The paper is outlined as follows. In Sec.~\ref{sec:Lag} we describe the general
SM extensions with CLFV couplings. Then we discuss
the relevant existing constraints on the CLFV couplings in Sec.~\ref{sec:cons}.
In Sec.~\ref{sec:sen} we present the sensitivity of neutrino trident production, future lepton
colliders, and a new state-of-the-art muonium-antimuonium
conversion experiment to the CLFV couplings of bileptons and compare it
with the existing low-energy constraints.
Our conclusions are drawn in Sec.~\ref{sec:Con}.

\section{General Lagrangian for charged lepton flavor violation}
\label{sec:Lag}

In this work we consider all possible\footnote{In principle one could extend the discussion to spin-2 fields.} scalar and vector bileptons with possible CLFV couplings~\cite{Cuypers:1996ia}. They are obtained by expanding the most general SM gauge invariant Lagrangian in terms of explicit leptonic fields.
The bileptons fall in two categories depending whether they carry lepton number $L$ or not.
The most general SM invariant Lagrangian of $\Delta L=0$ bileptons has four terms
\begin{align}
	\mathcal{L}_{\Delta L=0}&=
	y_1^{ij} H_{1\mu}^0 \bar{L}_i \gamma^\mu P_L L_j
+ y_1^{\prime ij} H_{1\mu}^{\prime 0} \bar{\ell}_i \gamma^\mu P_R \ell_j
+ \left( y_2^{ij} H_{2\alpha} \bar{L}_{i\alpha} P_R \ell_j + h.c. \right)
+ y_3^{ij} \bar{L}_i\gamma^\mu \vec{\sigma}\cdot \vec{H}_{3\mu} L_j
\nonumber\\&
=	
\left(
	y_1^{ij} H_{1\mu}^0 \bar{\ell}_i \gamma^\mu P_L \ell_j
+y_1^{ij} H_{1\mu}^0 \bar{\nu}_i \gamma^\mu P_L \nu_j
\right)
+ y_1^{\prime ij} H_{1\mu}^{\prime 0} \bar{\ell}_i \gamma^\mu P_R \ell_j
\nonumber \\&
+ \left( y_2^{ij} H_{2}^+ \bar{\nu}_i P_R \ell_j + y_2^{ij} H_{2}^0 \bar{\ell}_i P_R \ell_j + h.c. \right)
\nonumber \\&
+ \left(y_3^{ij}\sqrt{2}H_{3\mu}^-\bar{\ell}_i\gamma^\mu P_L \nu_j + y_3^{ij}\sqrt{2}H_{3\mu}^+\bar{\nu}_i\gamma^\mu P_L \ell_j - y_3^{ij}H_{3\mu}^0\bar{\ell}_i \gamma^\mu P_L \ell_j\right)\; ,
	\label{eq:delL0}
\end{align}
where $L_i=(\nu_i, \ell_i)$ denotes the left-handed SM lepton doublet with a flavor index $i$.
The subscript of the new bosonic fields, i.e.~1, 2 or 3, manifests their SU$(2)_L$ nature as singlet, doublet or triplet, respectively. The couplings $y_1^{(\prime)}$ and $y_3$ may arise from new gauge interactions with a LFV $Z^\prime$ or a SU(2)$_L$ triplet gauge boson and $y_2$ naturally appears in two Higgs doublet models with a complex neutral scalar $H_2^0=(h_2 + i a_2)/\sqrt{2}$. The couplings $y_1$, $y_1^\prime$, and $y_3$ are hermitian, while $y_2$ may take any values. Similarly, there are three different $\Delta L=2$ lepton bilinears
\begin{align}
	\mathcal{L}_{\Delta L=2}
	&= \lambda_1^{ij} \Delta_1^{++}\ell^T_i CP_R\ell_j
+  \lambda_2^{ij} \Delta_{2\mu\alpha}L^T_{i\beta} C\gamma^\mu P_R\ell_j \epsilon_{\alpha\beta}
-  {\lambda_3^{ij}\over \sqrt{2}} L_i^T C i\sigma_2\vec\sigma\cdot\vec\Delta_3 P_L L_j + h.c.
\nonumber\\
	&=\left( \lambda_1^{ij} \Delta_1^{++}\ell^T_i CP_R\ell_j + h.c.\right)
+ \left( \lambda_2^{ij} \Delta_{2\mu}^{++}\ell^T_i C\gamma^\mu P_R\ell_j - \lambda_2^{ij} \Delta_{2\mu}^{+}\nu^T_i C\gamma^\mu P_R\ell_j + h.c.\right) \nonumber \\
&- \left( -\lambda_3^{ij}\sqrt{2}\Delta_3^{+}\nu^T_i C P_L\ell_j -\lambda_3^{ij}\Delta_3^{++}\ell^T_i C P_L\ell_j + \lambda_3^{ij} \Delta_3^0 \nu^T_i C P_L \nu_j  + h.c. \right).\label{eq:delL2}
\end{align}
The neutral component of $\Delta_3$ only couples to the neutrino sector and thus it is irrelevant for our study of CLFV below. The couplings $\lambda_1$ and $\lambda_3$ are symmetric, while $\lambda_2$ may take arbitrary values. The $\Delta L=2$ coupling $\lambda_1$ naturally emerges in the Zee-Babu model which only couples to right-handed charged leptons~\cite{Zee:1985id,Babu:1988ki}, while $\lambda_3$ may come from the SU$(2)_L$ triplet field in the Type II Seesaw model which only interacts with left-handed charged leptons~\cite{Magg:1980ut,Schechter:1980gr,Lazarides:1980nt,Wetterich:1981bx,Mohapatra:1980yp}. The coupling $\lambda_2$ can arise after the breaking of a unified gauge model where the lepton doublet and the charge-conjugate of charged lepton singlet $\ell^c$ reside in the same multiplet. One example is an SU$(3)_c\times$SU$(3)_L\times$U$(1)_Y$ model~\cite{Pisano:1991ee}. See Ref.~\cite{Li:2018cod} for further details.
The non-zero elements of the above couplings, i.e. $y^{(\prime)ij}_{1,2,3}$, $\lambda^{ij}_{1,2,3}$, can lead to the presence of CLFV processes. Below we focus on the off-diagonal elements of the couplings which induce CLFV on-shell production of a bilepton $X$ with two different flavor charged leptons, although we present the general results for all possible bilepton interactions.

Models with new massive vector bosons generally require the introduction of a new Higgs boson with the exception of an Abelian vectorial symmetry where the mass of the gauge boson can be generated via the St\"uckelberg mechanism~\cite{Stueckelberg:1900zz}. This may lead to new contributions mediated by the components of the new Higgs boson. However, the processes which we are considering do not suffer from any theoretical problems like the violation of perturbative unitarity. Thus, to remain as model-independent as possible, we will assume that the contribution of the Higgs bosons to lepton flavor violating processes is negligible. This can be realized either by making them sufficiently heavy or by suppressing the off-diagonal couplings to leptons. We restrict ourselves to the Lagrangians in Eqs.~\eqref{eq:delL0} and \eqref{eq:delL2} for the rest of the paper. We do not take into account renormalization group corrections.

\section{Constraints}
\label{sec:cons}

In this section we summarize relevant constraints on the CLFV couplings from anomalous magnetic moments of leptons, muonium-antimuonium conversion, constraints from lepton flavor universality in leptonic $\tau$ decays, electroweak precision observables, and new leptonic non-standard neutrino interactions, and the existing collider searches. We only consider constraints which are relevant for the sensitivity study in Sec.~\ref{sec:sen}, e.g.~we do not consider $\mu\to e\gamma$, because it depends on a product of two independent couplings. See Ref.~\cite{Li:2018cod} for a study of other LFV processes mediated by bileptons.
Note that, although we give the analytical results for general coupling matrices, in this and the following sections we assume that there is no additional CP violation. Thus the couplings of $H_{1\mu}^{(\prime)}$, $H_{3\mu}$, and $\Delta_{1,3}$ are real and symmetric. For simplicity we further restrict the couplings for the electroweak doublets $H_2$ and $\Delta_{2\mu}$ to be symmetric in the numerical analysis.

\subsection{Anomalous magnetic moments}

\begin{figure}[tbp!]
\begin{center}
\includegraphics[scale=1,width=15cm]{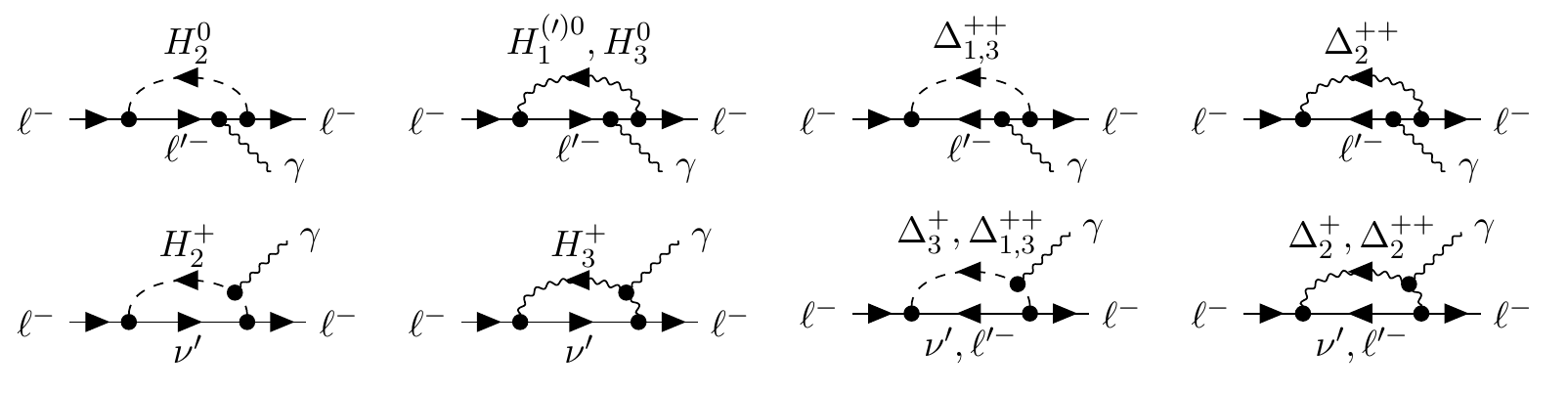}
\end{center}
\caption{Diagrams contributing to the anomalous magnetic moments by the Lagrangians in Eqs.~(\ref{eq:delL0}) and (\ref{eq:delL2}).}
\label{LFV:AMM}
\end{figure}

The muon magnetic dipole moment has $\sim 3.7\sigma$ discrepancy between the SM prediction~\cite{Blum:2018mom,Keshavarzi:2018mgv} and experimental measurements~\cite{Bennett:2006fi,Tanabashi:2018oca}
\begin{eqnarray}
\Delta a_\mu\equiv a_\mu^{\rm exp} - a_\mu^{\rm SM} = (2.74\pm 0.73)\times 10^{-9}.
\end{eqnarray}
For the electron $g-$2, Refs.~\cite{Parker:2018,Davoudiasl:2018fbb} recently presented a precise measurement with a $2.4\sigma$ discrepancy
\begin{eqnarray}
\Delta a_e\equiv a_e^{\rm exp} - a_e^{\rm SM} = (-0.88\pm 0.36)\times 10^{-12}.
\end{eqnarray}
Apparently, the muon (electron) AMM requires a positive (negative) new physics contribution to explain the discrepancy between the theoretical SM prediction and the experimental value.
The one-loop diagrams contributing to the AMM by the Lagrangians in Eqs.~(\ref{eq:delL0}) and (\ref{eq:delL2}) are shown in Fig.~\ref{LFV:AMM}.

Using the general formulas provided by Lavoura in Ref.~\cite{Lavoura:2003xp}, we find that the leading contributions of the vector bosons $H_1^{(\prime)0}$, $H_3$ and $\Delta_2$ to the anomalous magnetic moment of the lepton $\ell$ are respectively
\begin{align}
	\Delta a_{\ell}(H_1^{(\prime)0})
	&= \frac{(y_1^{(\prime)\dagger} y_1^{(\prime)})^{\ell\ell}}{12\pi^2} \frac{m_\ell^2}{m_{H_1^{(\prime)0}}^2} \geq 0 \ ,
	\\\nonumber
	\Delta a_\ell(H_3) &= \frac{(y_3^\dagger y_3)^{\ell\ell}}{12\pi^2} \frac{m_\ell^2}{m_{H_3^0}^2} -\frac{5(y_3^\dagger y_3)^{\ell\ell}}{24\pi^2} \frac{m_\ell^2}{m_{H_3^+}^2} \ ,
	\\\nonumber
	\Delta a_\ell(\Delta_2) &= -\frac{7(\lambda_2^\dagger \lambda_2)^{\ell\ell}}{24\pi^2} \frac{m_\ell^2}{m_{\Delta_2^{++}}^2}  -\frac{5(\lambda_2^\dagger \lambda_2)^{\ell\ell}}{48\pi^2} \frac{m_\ell^2}{m_{\Delta_2^+}^2}  \leq 0\; ,
\end{align}
to leading order in the charged lepton mass.
For the scalars $\Delta_{1,3}$ and $H_2$, the new AMMs are given by
\begin{align}
		\Delta a_{\ell}(\Delta_1) =& \frac{(\lambda_1^\dagger \lambda_1)^{\ell\ell}} {6\pi^2} \frac{m_\ell^2}{m_{\Delta_1^{++}}^{2}} \geq 0 \ ,\\
		\nonumber
		\Delta a_{\ell}(\Delta_3) =& \frac{(\lambda_3^\dagger \lambda_3)^{\ell\ell}} {6\pi^2} \left(\frac{m_\ell^2}{m_{\Delta_3^{++}}^{2}} + \frac{m_\ell^2}{8m_{\Delta_3^{+}}^{2}} \right) \geq 0 \ ,
		\\ \nonumber
		\Delta a_\ell(H_2) =&
	-\frac{(y_2^\dagger y_2+y_2y_2^\dagger)^{\ell\ell}}{96\pi^2} \left(\frac{m_\ell^2}{m_{h_2}^2}+ \frac{m_\ell^2}{m_{a_2}^2}  \right) + \frac{ (y_2^\dagger y_2)^{\ell\ell}}{96\pi^2} \frac{m_\ell^2}{m_{H_2^+}^2}
	\\\nonumber &
	+\sum_k\mathrm{Re}[y_2^{k\ell}y_2^{\ell k}]  \frac{m_k m_\ell}{16\pi^2} \left(\frac{\ln \left(\tfrac{m_k^2}{m_{h_2}^2}\right) +\tfrac32}{m_{h_2}^2} - \frac{\ln \left(\tfrac{m_k^2}{m_{a_2}^2}\right) +\tfrac32}{m_{a_2}^2} \right)\; ,
\end{align}
to leading order in the charged lepton masses. One can see that, apart from $H_2$ and $H_3$, each new contribution to the anomalous magnetic moment has a definite sign. For $H_2$ in the limit of degenerate scalar masses $m_{h_2}=m_{a_2}$, the anomalous magnetic moment is not enhanced proportional to the mass of the $\tau$ lepton in the loop and the contribution obtains a definite negative sign
\begin{align}
\Delta a_\ell(H_2)
& =
	-\frac{(y_2^\dagger y_2+y_2y_2^\dagger)^{\ell\ell}}{48\pi^2} \frac{m_\ell^2}{m_{h_2}^2}
	\leq 0 \ \ \ {\rm for} \ m_{h_2}=m_{a_2}\ll m_{H_2^+}\; ; \nonumber\\
\Delta a_\ell(H_2)
& =  -\frac{(y_2^\dagger y_2)^{\ell\ell}+ 2 (y_2y_2^\dagger)^{\ell\ell}}{96\pi^2} \frac{m_\ell^2}{m_{h_2}^2}
	\leq 0 \ \ \ {\rm for} \ m_{h_2}=m_{a_2}=m_{H_2^+}\;.
\end{align}
Similarly, if all scalars apart from the CP-even neutral scalar $h_2$ are decoupled, the contribution is negative.
In other extreme limits, such as $m_{a_2}\ll m_{h_2}, m_{H_2^+}$ or $m_{H_2^+}\ll m_{h_2}, m_{a_2}$, the anomalous magnetic moment may be positive. For the $H_3$ case, the contribution to the anomalous magnetic moment becomes negative in the limit of degenerate scalars $m_{H_3^0}=m_{H_3^+}$ and positive for $m_{H_3^0}\ll m_{H_3^+}$.

As a result, $H_1^{(\prime)0}$ and $\Delta_{1,3}$ can only explain the deviation in $a_\mu$, while $\Delta_2$ can explain $a_e$. The contributions from $H_{2,3}$ can have either sign and thus in principle address both anomalies. In this work we do not attempt to explain the deviations from the SM but rather derive a constraint on the LFV couplings described in Eqs.~(\ref{eq:delL0}) and (\ref{eq:delL2}). In order to derive a constraint, we demand that the
new physics contribution deviates from the experimental observation by at most $3\sigma$ for the electron and $4\sigma$ for the muon in order to account for the discrepancies in both measurements. The constraints from the AMMs are summarized in Table~\ref{tab:AMM}.

\begin{table}[tbp!]
\begin{center}
\begin{tabular}{|c|c|c|}
\hline
 & $a_e\; [3\sigma]$ & $a_\mu\; [4\sigma]$\\\hline
$H_1^{(\prime)0}$
& $|(y_1^{(\prime)\dagger} y_1^{(\prime)})^{ee}|< 9.1\times 10^{-5}\, m_{H_1^{(\prime)0}}^2 $
& $|(y_1^{(\prime)\dagger} y_1^{(\prime)})^{\mu\mu}| < 6.0\times 10^{-5}\ m_{H_1^{(\prime)0}}^2$\\\hline
$H_2$
& $|(y_2^\dagger y_2)^{ee}|< 1.8 (2.4)\times 10^{-3}\, m_{h_2}^2 $
& $|(y_2^\dagger y_2)^{\mu\mu}| < 3.8 (5.1)\times 10^{-6}\ m_{h_2}^2$\\\hline
$H_3$
& $|(y_3^\dagger y_3)^{ee}|< 9.1 (59)\times 10^{-5}\, m_{H_3^0}^2 $
& $|(y_3^\dagger y_3)^{\mu\mu}|< 6.0 (0.13)\times 10^{-5}\, m_{H_3^0}^2 $
\\\hline
$\Delta_1$
& $|(\lambda_{1}^\dagger \lambda_{1})^{ee}| < 4.5\times 10^{-5}\, m_{\Delta_{1}^{++}}^2$
& $|(\lambda_{1}^\dagger \lambda_{1})^{\mu\mu}| < 3.0\times 10^{-5}\ m_{\Delta_{1}^{++}}^2$ \\\hline
$\Delta_2$
& $|(\lambda_2^\dagger \lambda_2)^{ee}|< 2.5 (1.9)\times 10^{-4}\, m_{\Delta_2^{++}}^2 $
& $|(\lambda_2^\dagger \lambda_2)^{\mu\mu}| < 5.5 (4.0)\times 10^{-7}\ m_{\Delta_2^{++}}^2$ \\\hline
$\Delta_3$
& $|(\lambda_3^\dagger \lambda_3)^{ee}|< 4.5 (4.0)\times 10^{-5}\, m_{\Delta_3^{++}}^2 $
& $|(\lambda_3^\dagger \lambda_3)^{\mu\mu}| < 3.0 (2.7)\times 10^{-5}\ m_{\Delta_3^{++}}^2$
\\\hline
\end{tabular}
\end{center}
\caption{Constraints from AMM on the CLFV couplings in units of ${\rm GeV}^{-2}$. Here we assume all the CLFV couplings are real and symmetric, and $m_{h_2}=m_{a_2}$ for $H_2$. For $H_2$, $H_3$, $\Delta_2$ and $\Delta_3$, the values outside the brackets correspond to the assumption that the singly charged boson is decoupled, while values in the brackets are under the assumption that all components of the multiplet are degenerate.
}
\label{tab:AMM}
\end{table}

\subsection{Muonium-antimuonium conversion}

Muonium is the bound state of $\mu^+$ and $e^-$ and antimuonium is that of $\mu^-$ and $e^+$. If there is a mixing of muonium ($M=(\mu^+ e^-)$) and antimuonium ($\bar M=(\mu^- e^+)$), the lepton flavor conservation of electron and muon must be violated and thus it is a sensitive probe for CLFV.

The probability of muonium-antimuonium conversion has been firstly calculated in Refs.~\cite{Feinberg:1961zz,Feinberg:1961zza}. Following the discussions in Refs.~\cite{Feinberg:1961zz,Feinberg:1961zza,Matthias:1991fp,Horikawa:1995ae}, we use the density matrix formalism to calculate the probability of muonium to antimuonium conversion. In contrast to previous calculations~\cite{Horikawa:1995ae}, we include off-diagonal elements in the Hamiltonian $H_{M\bar M}$ which mediates muonium-antimuonium conversion, and expand to the first order in the interaction Hamiltonian $H_{M\bar M}$. This is generally a good approximation for $B\gtrsim 0.1 \ \mu$T assuming at most weak-scale interaction strength for $H_{M\bar M}$.

Muonium is described by the Hamiltonian
\begin{eqnarray}
H=H_0+H_{hf}+H_Z\; ,
\end{eqnarray}
where $H_0$ denotes the non-relativistic Hamiltonian for a hydrogen-like system, i.e.~a bound state of two particles via a Coulomb interaction. The hyperfine splitting of the $1s$ state is described by $H_{hf}=b \,\vec S_\mu\cdot\vec S_e$ with $b\simeq 1.85\times 10^{-5}$ eV~\cite{Mariam:1982bq,Klempt:1982ge}, where $\vec S_{e,\mu}$ are the spins of the electron and muon, respectively. Finally, $H_Z=-(\vec \mu_e+\vec \mu_\mu)\cdot \vec B$ describes the Zeeman effect with external magnetic field $\vec B$. The magnetic moments for electron and muon are defined as $\vec \mu_{e}=-g_e \mu_B\vec S_{e}$ and $\vec\mu_\mu=g_\mu \mu_B \vec S_\mu m_e/m_\mu$ with two $g$-factors $g_{e,\mu}\simeq2\,(1+\alpha/2\pi)$ and the Bohr magneton $\mu_B=e/2m_e$.

In the uncoupled basis $\ket{\uparrow\uparrow}$, $\ket{\uparrow\downarrow}$, $\ket{\downarrow\uparrow}$, $\ket{\downarrow\downarrow}$, the Hamiltonian can be expressed as
\begin{equation}
	H =E_0\; \mathbf{1}+ \frac{b}{2} \begin{pmatrix}
		\frac12+Y &&&\\
		    & -\frac12+X & 1 &\\
		    & 1 & -\frac12-X &\\
		    &&& \frac12-Y\\
	\end{pmatrix}\; ,
\end{equation}
in terms of the ground state energy $E_0=-\alpha^2 m_{red}/2$, the fine structure constant $\alpha$, the reduced mass of the two-body system $m_{red}=m_e m_\mu/(m_e+m_\mu)\simeq m_e$ and two functions
\begin{align}
	X&=\frac{\mu_B B}{b} \left(g_e +\frac{m_e}{m_\mu} g_\mu\right) \; , &
	Y&=\frac{\mu_B B}{b} \left(g_e -\frac{m_e}{m_\mu} g_\mu\right)\;,
\end{align}
which parameterize the Zeeman effect. The energy eigenstates with their eigenenergies are thus given by
\begin{align}
	\ket{\lambda_1^{(M)}} & = \ket{M; \uparrow\uparrow} &
	\lambda_1^{(M)} & = E_0 +\frac{b}{2}\left(\frac12+Y\right) \; ,\\
	\ket{\lambda_2^{(M)}} & = c \ket{M; \uparrow\downarrow} +  s \ket{M; \downarrow\uparrow}
			      &
	\lambda_2^{(M)} & = E_0 +\frac{b}{2}\left(-\frac12+\sqrt{1+X^2}\right) \; ,\\
\ket{\lambda_3^{(M)}} & = -s \ket{M; \uparrow\downarrow} +c \ket{M; \downarrow\uparrow}
			      &
	\lambda_3^{(M)} & = E_0 +\frac{b}{2}\left(-\frac12-\sqrt{1+X^2}\right) \; ,\\
\ket{\lambda_4^{(M)}} & = \ket{M; \downarrow\downarrow}
			      &
	\lambda_4^{(M)} & = E_0 +\frac{b}{2}\left(\frac12-Y\right) \; ,
\end{align}
and the mixing is described by
\begin{align}
	s & = \left(\frac{\sqrt{1+X^2}-X}{2\sqrt{1+X^2}}\right)^{1/2}
	\qquad\mathrm{and}\qquad
	c  = \left(\frac{\sqrt{1+X^2}+X}{2\sqrt{1+X^2}}\right)^{1/2}\;.
\end{align}
For a vanishing magnetic field $s=c=1/\sqrt{2}$ and thus $\ket{\lambda_{3}^{(M)}}$ becomes the singlet state, while $\ket{\lambda_{1,2,4}^{(M)}}$ form the triplet state. The corresponding expressions for antimuonium are obtained with the replacements
\begin{align}
	(X,Y) &\to (-X,-Y) \; , &
	s&\leftrightarrow c\;.
\end{align}

\begin{figure}[tbp!]
\begin{center}
\includegraphics[scale=1,width=12cm]{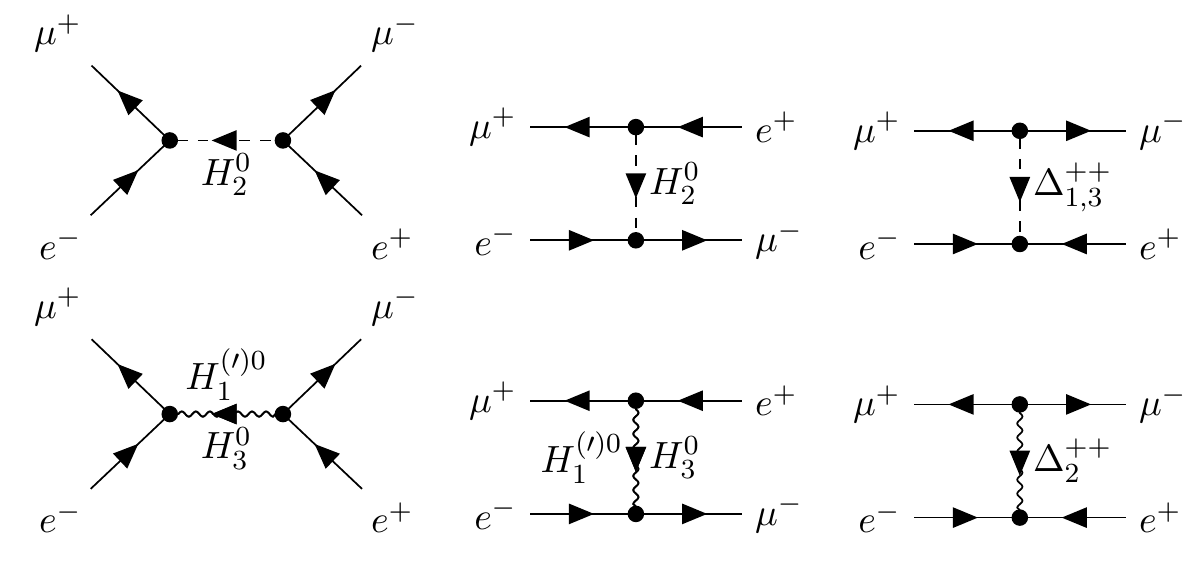}
\end{center}
\caption{Diagrams contributing to the muonium-antimuonium conversion by the Lagrangians in Eqs.~(\ref{eq:delL0}) and (\ref{eq:delL2}).}
\label{LFV:conversion}
\end{figure}

The interaction Hamiltonian $H_{M\bar M}$ inducing the muonium-antimuonium conversion may have different forms. We are particularly interested in the following vector and scalar interactions with equal and opposite chirality leptons
\begin{align}
 H_{LL(RR)}\equiv[\bar \mu \gamma_\rho P_{L(R)} e] [\bar\mu \gamma^\rho P_{L(R)} e]
&\qquad\to\qquad
\frac{2}{\pi a^3}\begin{pmatrix}
	1 &&&\\
	  & \frac{1}{\sqrt{1+X^2}} & \frac{X}{\sqrt{1+X^2}} &\\
	  & -\frac{X}{\sqrt{1+X^2}} & \frac{1}{\sqrt{1+X^2}} & \\
	  &&&1
\end{pmatrix}\;,
\label{HLLRR}
\\
 H_{LR}\equiv[\bar \mu \gamma_\rho P_L e] [\bar\mu \gamma^\rho P_R e]
&\qquad\to\qquad
 \frac{1}{\pi a^3} \begin{pmatrix}
	1 &&&\\
	  & 2-\frac{1}{\sqrt{1+X^2}} & \frac{X}{\sqrt{1+X^2}} &\\
	  & -\frac{X}{\sqrt{1+X^2}} & -2-\frac{1}{\sqrt{1+X^2}} & \\
	  &&&1
\end{pmatrix}\;,
\label{HLR}
\end{align}
\begin{align}
 H_{SLL(SRR)}\equiv[\bar \mu P_{L(R)} e] [\bar\mu  P_{L(R)} e]
&\qquad\to\qquad
 -\frac14 H_{LL(RR)}\;,
\label{HSLLRR}
\\
 H_{SLR}\equiv[\bar \mu  P_L e] [\bar\mu  P_R e]
&\qquad\to\qquad
 -\frac12 H_{LR}\;,
\end{align}
where the matrix representation on the right-hand side is in the basis $\braket{\lambda_i^{(\bar M)}|\hat H|\lambda_{j}^{(M)}}$ and $a$ denotes the Bohr radius $a=(1/\alpha) (m_e+m_\mu)/(m_e m_\mu) \simeq 1/\alpha m_e$. Note that the scalar interaction leads to the same Hamiltonians in matrix form as the vector interactions, with only a different overall factor.
The relevant contributions to muonium-antimuonium conversion from our Lagrangians are shown in Fig.~\ref{LFV:conversion} and result in
\begin{eqnarray}
\mathcal{L}(H_1^{(\prime)0}) &=& \frac{|y_1^{(\prime)\mu e}|^2}{2m_{H_1^{(\prime)0}}^2} \left[ \bar \mu \gamma^\mu P_{L(R)} e \right] \left[\bar\mu\gamma_\mu P_{L(R)} e \right]\; , \ \
	\mathcal{L}(H_3) = \frac{|y_3^{\mu e}|^2}{2m_{H_3^0}^2} \left[ \bar \mu \gamma^\mu P_{L} e \right] \left[\bar\mu\gamma_\mu P_{L} e \right]\; ,\\
\mathcal{L}(\Delta_{1,3}^{++}) &=& -\frac{\lambda_{1,3}^{ee} \lambda_{1,3}^{\mu\mu*}}{2m_{\Delta_{1,3}^{++}}^2} \left[ \bar \mu \gamma^\mu P_{R,L} e \right] \left[\bar\mu\gamma_\mu P_{R,L} e \right]\; , \ \
\mathcal{L}(\Delta_2) = - \frac{\lambda_2^{ee} \lambda_2^{\mu\mu*}}{m_{\Delta_2^{++}}^2} \left[\bar \mu \gamma^\mu P_L e\right] \left[\bar \mu  \gamma_\mu P_R e\right]\; ,\\
	\mathcal{L}(H_2)
 &=& \frac14 \left(\frac{1}{m_{h_2}^2}-\frac{1}{m_{a_2}^2}\right) \left[ (y_2^{\mu e})^2\, [\bar\mu P_R e ]^2 +  (y_2^{e \mu *})^2 \, [\bar \mu P_L e]^2 \right]\nonumber \\
	&-& \frac{y_2^{\mu e} y_2^{e\mu*}}{4} \left(\frac{1}{m_{h_2}^2} + \frac{1}{m_{a_2}^2} \right) [\bar\mu \gamma^\mu P_L e] [\bar \mu \gamma_\mu P_R e]\;.
\end{eqnarray}
The corresponding interaction Hamiltonians in the $\ket{\lambda_i}$ basis are
\begin{align}
H_{M\bar M}(H_1^{(\prime)0}) &= -\frac{|y_1^{(\prime)\mu e}|^2}{2m_{H_1^{(\prime)0}}^2} H_{LL(RR)}\; ,
             &
H_{M\bar M}(H_3) &=
-\frac{|y_3^{\mu e}|^2}{2m_{H_3^0}^2} H_{LL}\; ,\\
H_{M\bar M}(\Delta_{1,3}^{++}) &= \frac{\lambda_{1,3}^{ee}\lambda_{1,3}^{\mu\mu*}}{2m_{\Delta_{1,3}^{++}}^2} H_{RR,LL}\; ,
			       &
	H_{M\bar M}(\Delta_2) &= \frac{\lambda_2^{ee}\lambda_2^{\mu\mu*}}{m_{\Delta_2^{++}}^2} H_{LR}\;,
\end{align}
\begin{align}
H_{M\bar M}(H_2)
=C\, H_{LR} + \frac{A}{2} H_{LL}\; ,
\end{align}
with
\begin{align}\label{eq:AC}
	C&\equiv\frac{y_2^{\mu e}y_2^{e\mu*}}{4} \left(\frac{1}{m_{h_2}^2}+\frac{1}{m_{a_2}^2}\right)\;, &
	A&\equiv \frac{(y_2^{\mu e})^2 + (y_2^{e\mu*})^2}{8} \left(\frac{1}{m_{h_2}^2}-\frac{1}{m_{a_2}^2}\right)\;.
\end{align}
Note the non-trivial dependence on the magnetic field in case of the electroweak doublet scalar $H_2$. In the limit of degenerate scalar masses $m_{h_2}=m_{a_2}$, the effective Lagrangian and the interaction Hamiltonian induced by $H_2$ simplify to
\begin{align}
	\mathcal{L}(H_2)
 & =
	- \frac{y_2^{\mu e} y_2^{e\mu*}}{2\,m_{h_2}^2} [\bar\mu \gamma^\mu P_R e] [\bar \mu \gamma_\mu P_L e]\; ,
	&
H_{M\bar M}(H_2)&= \frac{y_2^{\mu e} y_2^{e\mu *}}{2m_{h_2}^2} H_{LR}\; ,
\end{align}
which are consistent with our results in Ref.~\cite{Li:2018cod}.

The probability to observe a $\mu^+$ decay instead of a $\mu^-$ decay starting from an unpolarized muonium is
\begin{align}
	P(B) & =\int_0^\infty dt\, \gamma\, e^{-\gamma t}\, \mathrm{tr}\left(e^{-iHt}\rho_0e^{iHt}\mathcal{P}_{\bar M}\right)
=\frac12 \sum_{i,j} \frac{\left|\braket{\lambda_i^{(\bar M)}|H_{M\bar M}|\lambda_{j}^{(M)}}\right|^2}{\gamma^2+(\lambda_i^{(\bar M)}-\lambda_j^{(M)})^2}\; ,
\end{align}
where $\gamma=G_F^2 m_\mu^5/192\pi^3$ is the muon decay rate, $\rho_0=\frac14 \sum_i \ket{\lambda_i^{(M)}}\bra{\lambda_i^{(M)}}$ is the density matrix of the initial state muonium and $\mathcal{P}_{\bar M}=\sum_i \ket{\lambda_i^{(\bar M)}}\bra{\lambda_i^{(\bar M)}}$ is the projection operator onto the final state antimuonium.
For the interaction Hamiltonians of interest, there is only mixing between the second and third state as seen from Eqs.~(\ref{HLLRR}) and (\ref{HLR}), and thus the probability can be explicitly written as
\begin{align}
	P(B)=&
	\frac{ \left|\braket{\lambda_1^{(\bar M)}| H_{M\bar M}|\lambda_1^{(M)}}\right|^2 + \left|\braket{\lambda_4^{(\bar M)}| H_{M\bar M}|\lambda_4^{(M)}}\right|^2}{2\gamma^2+ 2b^2 Y^2}
\\\nonumber &
+\frac{\left|\braket{\lambda_2^{(\bar M)}|H_{M\bar M}|\lambda_2^{(M)}}\right|^2+ \left|\braket{\lambda_3^{(\bar M)}|H_{M\bar M}|\lambda_3^{(M)}}\right|^2}{2\gamma^2}
\\\nonumber &
	+ \frac{\left|\braket{\lambda_2^{(\bar M)}|H_{M\bar M}|\lambda_3^{(M)}}\right|^2 +\left|\braket{\lambda_3^{(\bar M)}|H_{M\bar M}|\lambda_2^{(M)}}\right|^2}{2\gamma^2 +2b^2 (1+X^2) }\; ,
	\label{eq:mmbarprob}
\end{align}
and in particular for vanishing magnetic field, $B=0$, we find
\begin{align}
	P(0)&=
	\frac{\sum_i\left|\braket{\lambda_i^{(\bar M)}|H_{M\bar M}|\lambda_i^{(M)}}\right|^2}{2\gamma^2}\;.
\end{align}
We compared our result with the analytic expression in Ref.~\cite{Horikawa:1995ae} and the numerical values in Table II of Ref.~\cite{Willmann:1998gd} and found good agreement numerically, although the contributions from the off-diagonal entries in the interaction Hamiltonians $H_{M\bar M}$ were not included in Ref.~\cite{Horikawa:1995ae}. These additional contributions vanish given no external magnetic field $B$ and are generally subdominant at finite external magnetic field. They are suppressed by the factor of $\gamma^2/(\gamma^2+b^2(1+X^2))$ compared with the dominant contribution, because the weak decay rate $\gamma$ is much smaller than the hyperfine splitting and the Zeeman effect, i.e.~$\gamma\ll b,bX,bY$.

Typically there are magnetic fields in the experimental setup. They suppress the conversion probability, because the degeneracy of the energy levels in $M$ and $\bar M$ is lifted. For the Hamiltonians with same chirality vector currents $H_{LL(RR)}$ and opposite chirality vector currents $H_{LR}$, the suppression factors of the probability at a finite magnetic field $B$ are
\begin{align}
	S_{XX}(B)\equiv \frac{P_{LL,RR}(B)}{P_{LL,RR}(0)} & =\frac12\left[ \frac{\gamma^2 }{\gamma^2+  b^2 (1+X^2)} \frac{X^2}{1+X^2} +  \frac{1}{1+X^2}  + \frac{\gamma^2}{\gamma^2+ b^2 Y^2} \right]\;,
	\\
	S_{LR}(B)\equiv \frac{P_{LR}(B)}{P_{LR}(0)} & =\frac{1}{6}\left[ \frac{\gamma^2 }{\gamma^2+  b^2 Y^2} +\frac{X^2}{1+X^2} \frac{\gamma^2}{\gamma^2+b^2(1+X^2)}  + \frac{4X^2+5}{1+X^2} \right] \;,
\end{align}
respectively. In particular, we obtain the numerical values $S_{XX}(0.1 \ {\rm T})=0.36$ and $S_{LR}(0.1 \ {\rm T})=0.79$ for a magnetic field $B=0.1$ T. Our values are $O(1-2)\%$ larger than the results in Refs.~\cite{Horikawa:1995ae,Willmann:1998gd} due to the inclusion of additional off-diagonal entries in the Hamiltonian $\braket{\lambda_i^{(\bar M)}|H_{M\bar M}|\lambda_j^{(M)}}$ with $(i,j)=(2,3)$ or $(3,2)$, but we agree with the overall magnitude of the suppression factor.

For the Lagrangians described in Eqs.~(\ref{eq:delL0}) and (\ref{eq:delL2}), we obtain their probabilities as follows
\begin{align}
	P(H_1^{(\prime)0})& =\frac{2|y_1^{(\prime)\mu e}|^4}{ \pi^2 a^6 \gamma^2 m_{H_1^{(\prime)0}}^4}  S_{XX}(B)\; ,
			 &
P(H_3)& =\frac{2|y_3^{\mu e}|^4}{ \pi^2 a^6 \gamma^2 m_{H_3^0}^4} S_{XX}(B)\; ,\\
P(\Delta_{1,3}^{++})& =\frac{2|\lambda_{1,3}^{ee}\lambda_{1,3}^{\mu\mu*}|^2}{ \pi^2 a^6 \gamma^2 m_{\Delta_{1,3}^{++}}^4} S_{XX}(B)\; ,
			    &
	P(\Delta_2)& =\frac{6|\lambda_2^{ee}\lambda_2^{\mu\mu*}|^2}{\pi^2 a^6 \gamma^2m_{\Delta_2^{++}}^4}S_{LR}(B)\; ,
\end{align}
\begin{align}
	P(H_2) & = \frac{1}{\pi^2 a^6 \gamma^2}\left[
		4 |C|^2 + \frac{\left|A-C\right|^2}{1+X^2}
		+\frac{\gamma^2\,\left|A+C\right|^2 }{\gamma^2+b^2 Y^2}
		+\frac{X^2}{1+X^2} \frac{\gamma^2\,\left(|A|^2+|C|^2\right) }{\gamma^2+b^2(1+X^2)}
	\right]\; ,
\end{align}
where $A$ and $C$ are defined in Eq.~(\ref{eq:AC}).
Note the non-trivial dependence of $P(H_2)$ on the magnetic field.
For real symmetric Yukawa couplings, the probability for $H_2$ can be written as
\begin{align}
	P(H_2) & = \frac{|y_2^{\mu e}|^4}{4\pi^2 a^6 \gamma^2}\Big[
		 \left(\frac{1}{m_{h_2}^2}+\frac{1}{m_{a_2}^2}\right)^2
		+ \frac{1}{m_{a_2}^4}\frac{1}{1+X^2}
+		\frac{1}{m_{h_2}^4}\frac{\gamma^2 }{\gamma^2+b^2 Y^2}
	\\\nonumber&	
		+\frac12\left(\frac{1}{m_{h_2}^4} + \frac{1}{m_{a_2}^4}\right)\frac{X^2}{1+X^2} \frac{\gamma^2 }{\gamma^2+b^2(1+X^2)}
	\Big]\; ,
\end{align}
which simplifies to
\begin{align}
	P(H_2) & = \begin{cases}
		{3|y_2^{e\mu}|^4\over 2\pi^2\gamma^2 a^6 m_{h_2}^4}S_{LR}(B) &
		\ \ {\rm for} \ m_{h_2}=m_{a_2}\\
	   \frac{\left|y_2^{e\mu}\right|^4}{2\pi^2\gamma^2a^6m_{h_2}^4} S_{h_2}(B) &
	\ \ {\rm for} \ m_{h_2}\ll m_{a_2}
\end{cases}\;, \\
S_{h_2}(B)&=
\frac12\left[1+\frac{\gamma^2}{\gamma^2+b^2 Y^2} + \frac12 \frac{X^2}{1+X^2} \frac{\gamma^2}{\gamma^2+b^2(1+X^2)}\right] &
\mathrm{with}\;\;S_{h_2}(0.1\,\mathrm{T})=0.5\; .
\end{align}
The search for muonium-antimuonium conversion at the Paul Scherrer Institut (PSI) placed a constraint on the probability to observe the decay of the muon in antimuonium decay instead of the decay of the antimuon in muonium with a magnetic field of $B=0.1$ T, that is $P(B=0.1 \ {\rm T}) \leq 8.3\times 10^{-11}$~\cite{Willmann:1998gd}. This bound can be used to obtain the constraints on the CLFV couplings of the bileptons which we summarize in Table~\ref{tab:muonium}.

\begin{table}[tb!]
\begin{center}
\begin{tabular}{|c|c|}
\hline
 & $\mu^+ e^-\to \mu^- e^+$ \\\hline
$H_1^{(\prime)0}$
& $|y_1^{(\prime)e\mu}|^2< 2.0\times 10^{-7}\, m_{H_1^{(\prime)0}}^2 $\\\hline
$H_2$
& $|y_2^{e\mu}|^2< 1.6 \ (3.4)\times 10^{-7}\, m_{h_2}^2$\\\hline
$H_3$
& $|y_3^{e\mu}|^2< 2.0\times 10^{-7}\, m_{H_3^0}^2 $
\\\hline
$\Delta_1$
& $|\lambda_{1}^{ee}\lambda_{1}^{\mu\mu}|< 2.0\times 10^{-7}\, m_{\Delta_{1}^{++}}^2 $\\\hline
$\Delta_2$
& $|\lambda_2^{ee}\lambda_2^{\mu\mu}|< 7.8\times 10^{-8}\, m_{\Delta_2^{++}}^2 $\\\hline
$\Delta_3$
& $|\lambda_{3}^{ee}\lambda_{3}^{\mu\mu}|< 2.0\times 10^{-7}\, m_{\Delta_{3}^{++}}^2 $
\\\hline
\end{tabular}
\end{center}
\caption{Constraints from muonium-antimuonium conversion on the CLFV couplings in units of ${\rm GeV}^{-2}$. Here we assume all the CLFV couplings are real and symmetric. For $H_2$, the value outside (inside) the brackets is for the extreme case $m_{h_2}=m_{a_2}$ ($m_{h_2}\ll m_{a_2}$).
}
\label{tab:muonium}
\end{table}

\subsection{Lepton flavor universality}

The interactions of leptons with neutrinos lead to new contributions to effective operators with two leptons and two neutrinos. In the absence of light right-handed neutrinos, there are only two types of effective operators
\begin{align}\label{eq:NSI}
	\mathcal{L} & =
	- 2\sqrt{2} G_F[\bar \nu_i\gamma_\mu P_L \nu_j]  [\bar \ell_k\gamma^\mu \left( g_{LL}^{ijkl} P_L + g_{LR}^{ijkl}P_R\right) \ell_l]\;.
\end{align}
Both Wilson coefficients in Eq.~(\ref{eq:NSI}) are generated in the SM from the exchange of $W$ and $Z$ bosons and can be expressed in terms of the weak mixing angle $\theta_W$ as
\begin{align}
	g_{LL,SM}^{ijkl} & = \left(-\frac12+\sin^2\theta_W\right) \delta_{ij}\delta_{kl} + \delta_{il}\delta_{jk}
	\qquad\mathrm{and}\qquad
	g_{LR,SM}^{ijkl}  = \sin^2\theta_W \delta_{ij}\delta_{kl}\;.
\end{align}
The second term in the expression for $g_{LL,SM}$ originates from $W$ boson exchange, while the other ones are due to $Z$ boson exchange.
The new physics contributions to the two different sets of Wilson coefficients are given by
\begin{align}
	g_{LL,NP}^{ijkl}
	& =-\frac{y_1^{ij}y_1^{kl} }{2\sqrt{2}G_F m_{H_1^0}^2}
	-\frac{y_3^{kj}y_3^{il}}{\sqrt{2} G_F m_{H_3^+}^2}
	-\frac{\lambda_3^{jl} \lambda_3^{ik*}}{2\sqrt{2}G_F m_{\Delta_3^+}^2}\; ,
\label{gLL}
\\
g_{LR,NP}^{ijkl}
& =\frac{y_2^{il}y_2^{jk*}}{4\sqrt{2} G_Fm_{H_2^+}^2}
-\frac{\lambda_2^{il} \lambda_2^{jk*}}{2\sqrt{2}G_F m_{\Delta_2^+}^2}
\label{gLR}
\end{align}
and thus lepton flavor universality in lepton decays provides an interesting probe to the CLFV interactions for the bileptons. The relevant decay width for $\ell_1\to \ell_2\nu_1\bar\nu_2$ is~\cite{Kinoshita:1958ru,Marciano:1988vm,Ferroglia:2013dga,Fael:2013pja,Pich:2013lsa}
\begin{equation}\label{eq:muondecay}
	\Gamma(\ell_1\to \ell_2 \nu_1 \bar \nu_2(\gamma)) = \frac{G_F^2 m_\ell^5}{192\pi^3}\, |g_{LL}^{1221}|^2\, F\left(\frac{m_{\ell^\prime}^2}{m_\ell^2}\right)
	R_W
	R_\gamma
\end{equation}
in terms of the function $F(x)=1-8x+8x^3-x^4-12x^2\ln x$. The corrections due to the $W$ boson propagator and radiative corrections are respectively
\begin{align}
	R_W & =
	1+\frac35\frac{m_1^2}{m_W^2}+\frac95 \frac{m_{2}^2}{m_W^2}\; , &
	R_\gamma & =	1+\frac{\alpha(m_1)}{2\pi}\left(\frac{25}{4}-\pi^2\right)\; ,
\end{align}
where $m_{1,2}$ is the mass of the charged lepton $\ell_{1,2}$ and $\alpha(\mu)$ is the running fine structure constant at scale $\mu$ with the result at one-loop order as
\begin{equation}
\alpha(m_\ell)^{-1} = \alpha^{-1}-\frac{2}{3\pi}\ln\left(\frac{m_\ell}{m_e}\right)+\frac{1}{6\pi}\;.
\end{equation}
The most sensitive probes of LFU are the ratios
\begin{align}
	R_{\mu e} & = \frac{\Gamma(\tau\to \nu_\tau \mu \bar\nu_\mu)}{\Gamma(\tau\to \nu_\tau e \bar\nu_e)}  \simeq R_{\mu e}^{\rm SM} \left(1+2 \mathrm{Re}( g_{LL,NP}^{\tau \mu\mu \tau}-g_{LL,NP}^{\tau ee \tau})\right)	\; ,
	\\
	R_{\tau\mu} & = \frac{\Gamma(\tau\to \nu_\tau e \bar\nu_e)}{\Gamma(\mu\to \nu_\mu e \bar\nu_e)}
	\simeq R_{\tau\mu}^{\rm SM} \left(
	1+2 \mathrm{Re}(g_{LL,NP}^{\tau ee \tau}-g_{LL,NP}^{\mu ee \mu})
\right)\;,
\end{align}
which we expanded to leading order in the new physics contribution which interferes with the SM. Decays to other neutrino flavors which do not interfere with the SM lead to
\begin{align}
	R_{\mu e} & = R_{\mu e}^{\rm SM} \left\{1
		+ \sum_{\alpha,\beta}{}^\prime \left(\left|g_{LL,NP}^{\alpha\beta\mu\tau}\right|^2 - \left|g_{LL,NP}^{\alpha\beta\tau e}\right|^2\right)
	+ \sum_{\alpha,\beta} \left(\left|g_{LR,NP}^{\alpha\beta\mu\tau}\right|^2 - \left|g_{LR,NP}^{\alpha\beta\tau e}\right|^2 \right) \right\}\; ,\\
	R_{\tau\mu} & = R_{\tau\mu}^{\rm SM} \left\{1
		+ \sum_{\alpha,\beta}{}^\prime \left(\left|g_{LL,NP}^{\alpha\beta e\tau}\right|^2 - \left|g_{LL,NP}^{\alpha\beta e\mu}\right|^2\right)
	+ \sum_{\alpha,\beta} \left(\left|g_{LR,NP}^{\alpha\beta e\tau}\right|^2 - \left|g_{LR,NP}^{\alpha\beta e\mu}\right|^2 \right) \right\}
	\;,
\end{align}
where the prime indicates that we only sum over contributions which do not interfere.
Taking into account both the experimental errors and the uncertainties in the SM prediction, the current experimental values and errors\footnote{We use the PDG~\cite{Tanabashi:2018oca} values and uncertainties in addition to the parameters given in Table~\ref{tab:ewo_input}.
}
are
\begin{align}
	\frac{R_{\mu e}^{\rm exp}}{R_{\mu e}^{\rm SM}}&
	=1.0034\pm0.0032
	\qquad\mathrm{and}\qquad
	\frac{R_{\tau\mu}^{\rm exp}}{R_{\tau\mu}^{\rm SM}}
	=1.0022\pm0.0028	\;.
\end{align}
Thus, at the $2\sigma$ level, the relevant constraints are
	\begin{align}
	-0.0015 &< \mathrm{Re}(g_{LL,NP}^{\tau \mu\mu \tau}-g_{LL,NP}^{\tau ee \tau}) <0.0049\; ,
	\\
	-0.0017 &<\mathrm{Re}(g_{LL,NP}^{\tau ee \tau}-g_{LL,NP}^{\mu ee \mu})<0.0039
\end{align}%
for the contributions interfering with the SM.
The constraints on the non-interfering contributions are
\begin{align}
	-0.0030 &<\sum_{\alpha,\beta}{}^\prime \left(\left|g_{LL,NP}^{\alpha\beta\mu\tau}\right|^2 - \left|g_{LL,NP}^{\alpha\beta\tau e}\right|^2\right)
	+ \sum_{\alpha,\beta} \left(\left|g_{LR,NP}^{\alpha\beta\mu\tau}\right|^2 - \left|g_{LR,NP}^{\alpha\beta\tau e}\right|^2 \right) <0.0098
	\;,
	\\
	-0.0034 &<\sum_{\alpha,\beta}{}^\prime \left(\left|g_{LL,NP}^{\alpha\beta e\tau}\right|^2 - \left|g_{LL,NP}^{\alpha\beta e\mu}\right|^2\right)
	+ \sum_{\alpha,\beta} \left(\left|g_{LR,NP}^{\alpha\beta e\tau}\right|^2 - \left|g_{LR,NP}^{\alpha\beta e\mu}\right|^2 \right) <0.0078
\end{align}
and thus generally weaker. In the collider analysis we always only consider a single coupling. Thus a single operator dominates and these bounds can be translated to constraints on the CLFV couplings. In Table~\ref{tab:LFU} we collect the relevant constraints for the sensitivity study in Sec.~\ref{sec:sen}. The constraints on $H_1^0$ and the charged component of $\Delta_3$ are due to the interference with the SM contribution, while others are derived from the non-interfering part.
\begin{table}
\begin{tabular}{|c|c|c|}\hline
		 & $R_{\mu e}$ & $R_{\tau\mu}$ \\\hline
	 \multirow{2}{*}{$H_1^0$}	& $(y_1^{\mu\tau})^2 < 4.9\times 10^{-8} m_{H_1^0}^2$
	& $(y_1^{e\tau})^2 < 5.6\times 10^{-8} m_{H_1^0}^2$ \\
    & $(y_1^{e\tau})^2 < 1.6\times 10^{-7} m_{H_1^0}^2$ & $(y_1^{e\mu})^2 < 1.3\times 10^{-7} m_{H_1^0}^2$
		\\\hline
    \multirow{2}{*}{$H_2$}
    & $|y_2^{\mu\tau}|^2< 6.5\times 10^{-6} m_{H_2^+}^2$ & $|y_2^{e\tau}|^2< 5.8\times 10^{-6} m_{H_2^+}^2$ \\
    & $|y_2^{e\tau}|^2< 3.6\times 10^{-6} m_{H_2^+}^2$ & $|y_2^{e\mu}|^2< 3.8\times 10^{-6} m_{H_2^+}^2$
    \\ \hline
    \multirow{2}{*}{$H_3$}
    & $|y_3^{\mu\tau}|^2< 1.6\times 10^{-6} m_{H_3^+}^2$ & $|y_3^{e\tau}|^2< 1.5\times 10^{-6} m_{H_3^+}^2$ \\
    & $|y_3^{e\tau}|^2< 9.0\times 10^{-7} m_{H_3^+}^2$ & $|y_3^{e\mu}|^2< 9.6\times 10^{-7} m_{H_3^+}^2$
    \\ \hline
    \multirow{2}{*}{$\Delta_2$}
    & $|y_2^{\mu\tau}|^2< 3.3\times 10^{-6} m_{\Delta_2^+}^2$ & $|y_2^{e\tau}|^2< 2.9\times 10^{-6} m_{\Delta_2^+}^2$ \\
    & $|y_2^{e\tau}|^2< 1.8\times 10^{-6} m_{\Delta_2^+}^2$ & $|y_2^{e\mu}|^2< 1.9\times 10^{-6} m_{\Delta_2^+}^2$
    \\ \hline
    \multirow{2}{*}{$\Delta_3$} & $|\lambda_3^{\mu\tau}|^2 < 4.9\times 10^{-8} m_{\Delta_3^+}^2$ & $|\lambda_3^{e\tau}|^2 < 5.6\times 10^{-8} m_{\Delta_3^+}^2$ \\
    & $|\lambda_3^{e\tau}|^2 < 1.6\times 10^{-8} m_{\Delta_3^+}^2$ & $|\lambda_3^{e\mu}|^2 < 1.3\times 10^{-7} m_{\Delta_3^+}^2$
    \\\hline
\end{tabular}
\caption{Relevant constraints from lepton flavor universality of leptonic $\tau$ decays on the CLFV couplings in units of GeV$^{-2}$. The constraints on $H_1^0$ and $\Delta_3$ are due to the interference with the SM and thus stronger.}
\label{tab:LFU}
\end{table}

\subsection{New contribution to muon decay}
As the Fermi constant is extracted from muon decay, a new contribution to muon decay via the operators in Eqs.~\eqref{eq:NSI} leads to an effective shift of the Fermi constant. We denote the Fermi constant extracted from muon decay by $G_{F,\mu}$ and the SM Fermi constant as $G_F$. The rate of muon decay to an electron and two neutrinos is given by Eq.~\eqref{eq:muondecay}. As the final neutrino flavors in muon decay are not measured there may be new contributions which do not interfere with the SM. Thus we find for the Fermi constant extracted in muon decay
\begin{equation}
	G_{F,\mu}^2  = G_F^2\left( |1+g_{LL,NP}^{\mu ee\mu}|^2
		+ \sum_{\alpha,\beta}{}^\prime |g_{LL,NP}^{\alpha\beta e\mu}|^2
	+ \sum_{\alpha,\beta} |g_{LR,NP}^{\alpha\beta e\mu}|^2\right)\; ,
\end{equation}
where the prime on the summation sign indicates that we are not summing over the interfering component with $(\alpha,\beta)=(\mu,e)$. Taking $G_{F,\mu}$ as input, we find to leading order the modification of the Fermi constant in terms of different Wilson coefficients
\begin{equation}
G_{F}  = G_{F,\mu} \left( 1 + \delta G_F \right)\; ,
\qquad \delta G_F \equiv - \mathrm{Re}(g_{LL,NP}^{\mu ee\mu}) -\frac12 \sum_{\alpha,\beta}{}^\prime |g_{LL,NP}^{\alpha\beta e\mu}|^2 - \frac12\sum_{\alpha,\beta} |g_{LR,NP}^{\alpha\beta e\mu}|^2\;.
\end{equation}
This change of the Fermi constant leads to the modifications of other observables, in particular the weak mixing angle, the $W$ boson mass and the unitarity of the CKM matrix.
\begin{table}
	\begin{tabular}{|c|c|}\hline
		input & value\\\hline
		$m_Z$ [GeV]  & $91.1876\pm0.0021$\cite{Tanabashi:2018oca}\\\hline
		$G_{F,\mu}$ [GeV$^{-2}$] & $1.1663787(6)\times 10^{-5}$\cite{Tanabashi:2018oca}\\\hline
	$\alpha^{-1}$ & $137.035999046(27)$\cite{Parker:2018} \\\hline
	\end{tabular}
\caption{Input parameters for electroweak observables.}
\label{tab:ewo_input}
\end{table}
\textbf{Weak mixing angle.}
In the SM the weak mixing angle is given by~\cite{Brivio:2017vri,Tanabashi:2018oca}
\begin{equation}
	s_W^2\equiv \sin^2 \theta_W = \frac12\left[1-\sqrt{1-\frac{4\pi\alpha}{\sqrt{2} G_F  m_Z^2(1-\Delta r)}}\,\right]\;,
\end{equation}	
where $\Delta r$ parameterizes the loop corrections in the SM. It depends both on the top quark and Higgs masses and is currently given by
$\Delta r=0.03672\mp0.00017\pm0.00008$~\cite{Tanabashi:2018oca}, where the first uncertainty is from the top quark mass and the second from $\alpha(m_Z)$.
The shift in the Fermi constant leads to a shift in the weak mixing angle
\begin{equation}
\frac{\delta s_W^2}{s_W^2} = {s_{W,exp}^2 - s_W^2\over s_W^2}
= \frac{c_W^2}{s_W^2-c_W^2}\delta G_F \;.
\end{equation}
Using the Fermi constant extracted in muon decay $G_{F,\mu}$ together with $m_Z$ and $\alpha$ as input parameters, which are given in Table~\ref{tab:ewo_input}, we find numerically
	$s_W^2 = 0.22344\pm0.00006$.
A comparison with the on-shell value $s_{W,exp}^2 = 0.22343\pm0.00007$ in PDG~\cite{Tanabashi:2018oca} leads to a constraint on the shift in the Fermi constant
\begin{equation}\label{eq:weakmixing}
	-0.00056<\delta G_F < 0.00062
\end{equation}
at the $2\sigma$ level.
	
\textbf{$W$ boson mass.}
The $W$ boson mass has been measured very precisely to be $m_{W,\rm exp}=(80.379\pm0.012)$ GeV~\cite{Tanabashi:2018oca} and the current SM prediction is $m_{W,\rm SM}=(80.363\pm 0.020)$ GeV~\cite{Tanabashi:2018oca}.
Adding the errors in quadrature, we obtain
\begin{equation}\label{eq:mWexp}
	\frac{m_{W,\rm exp}^2}{m_{W,\rm SM}^2} = 1.00040 \pm0.00058\;,
\end{equation}
i.e. the experimental measurement of the $W$ boson mass is consistent with the SM prediction at $1\sigma$.
The SM prediction of the $W$ boson mass depends on the value of the Fermi constant. In the on-shell scheme, it is given by~\cite{Tanabashi:2018oca}
\begin{equation}
m_{W}^2  = \frac{\pi\alpha}{\sqrt{2}G_F \sin^2\theta_W (1-\Delta r)}\;.
\end{equation}
Thus a new contribution to muon decay leads to an effective change in the SM prediction of the $W$ boson mass, even though there are no direct contributions to the $W$ boson mass. In the $(G_F,m_Z,\alpha)$ scheme we find to leading order for the shift in the $W$ boson mass
\begin{equation}
\frac{\delta m_W^2}{m_W^2} \equiv \frac{m_{W,exp}^2}{m_W^2}-1
= - \left(\delta G_F + \frac{\delta s_W^2}{s_W^2}\right) = \frac{s_W^2}{c_W^2-s_W^2}\delta G_F\;.
\end{equation}
Thus, the experimental result \eqref{eq:mWexp} translates into a constraint on the Fermi constant at $2\sigma$
\begin{equation}\label{eq:mW}
-0.00206<\delta G_F<0.00423\; .
\end{equation}

\textbf{Unitarity of the CKM matrix.}
The unitarity of the Cabibbo-Kobayashi-Maskawa (CKM) mixing matrix has been measured very precisely. In particularly the relation for the first row reads~\cite{Tanabashi:2018oca}
\begin{equation}
	|V_{ud}|^2+|V_{us}|^2+|V_{ub}|^2=0.9994\pm0.0005\;.
\end{equation}
As the CKM matrix elements $V_{uq}$ are measured in leptonic meson decays $M^+\to \ell^+ \nu$, $V_{ud}$ also in beta decay, a modification of the Fermi constant extracted from muon decay leads to a violation of unitarity for the measured CKM matrix elements $V^M_{uq}$
\begin{equation}
	|V^M_{ud}|^2+|V^M_{us}|^2+|V^M_{ub}|^2= \frac{G_{F}^2}{G_{F,\mu}^2}=1+2\delta G_F
	\;.
\end{equation}
This can be translated in a constraint on the Fermi constant at $2\sigma$
\begin{equation}\label{eq:CKM}
-0.0008<\delta G_F<0.0004\; .
\end{equation}

\textbf{Combined constraint on the Fermi constant.}
Taking the most stringent constraints on the Fermi constant from Eq.~\eqref{eq:weakmixing}, Eq.~\eqref{eq:mW}, and Eq.~\eqref{eq:CKM} we obtain
\begin{equation}
-0.00056<\delta G_F<0.0004\; ,
\end{equation}
where the lower bound comes from the weak mixing angle and the upper bound from CKM unitarity. It translates into a constraint on the Wilson coefficients
\begin{align}
-0.0004<\mathrm{Re}(g_{LL,NP}^{\mu ee\mu})&<0.00056\; , &
 \left|g_{LL,NP}^{\alpha\beta e\mu}\right|, \left|g_{LR,NP}^{\alpha\beta e\mu}\right| &< 0.033\;.
\end{align}
The constraints for the different bileptons are collected in the second column of Table~\ref{tab:GF}.

\begin{table}[bt]
		\begin{tabular}{|c|c|c|c|}\hline
			 & Fermi constant & neutrino $\mu-e$ & neutrino $\tau - e$
\\\hline
	$H_1^0$	
	& $(y_1^{e\mu})^2 < 1.3\times10^{-8} m_{H_1^0}^2$
	& $(y_1^{e\mu})^2 < 1.6\times10^{-6} m_{H_1^0}^2$
	&
	\\\hline
		$H_2$
		& $|y_2^{e\mu}|^2 < 2.2\times 10^{-6} m_{H_2^+}^2$
	& $|y_2^{e\mu}|^2 < 3.3\times10^{-6} m_{H_2^+}^2$
	& $|y_2^{\tau e}|^2 < 5.7\times10^{-5} m_{H_2^+}^2$
	
		\\ \hline
		$H_3$
		& $|y_3^{e\mu}|^2 < 5.4\times 10^{-7} m_{H_3^+}^2$
   		& $(y_3^{e\mu})^2 < 8.2\times 10^{-7} m_{H_3^+}^2$
		& $(y_3^{e\tau})^2 < 5.3\times 10^{-6} m_{H_3^+}^2$

		\\ \hline
		$\Delta_2$
		& $|\lambda_2^{e\mu}|^2 < 1.1\times 10^{-6} m_{\Delta_2^+}^2$
    & $|\lambda_2^{e\mu}|^2 < 1.6\times 10^{-6} m_{\Delta_2^+}^2$
    & $|\lambda_2^{e\tau}|^2 < 1.6\times 10^{-5} m_{\Delta_2^+}^2$

		\\ \hline
    $\Delta_3$
    & $|\lambda_3^{e\mu}|^2 < 1.3\times 10^{-8} m_{\Delta_3^+}^2$
    & $|\lambda_3^{e\mu}|^2 < 2.0\times 10^{-6} m_{\Delta_3^+}^2$
    & $|\lambda_3^{e\tau}|^2 < 1.1\times 10^{-5} m_{\Delta_3^+}^2$

    \\\hline
\end{tabular}
\caption{Constraints from the Fermi constant and neutrino physics on the CLFV couplings in units of GeV$^{-2}$.}
\label{tab:GF}
\end{table}


\subsection{Non-standard neutrino interactions}
Several constraints have been derived for the Wilson coefficients in Eq.~\eqref{eq:NSI} from neutrino-electron interactions
\begin{align}\label{eq:NSI1}
	|g_{LL,NP}^{\mu\mu ee}| & < 0.030~\text{\cite{Davidson:2003ha,Barranco:2007ej}}\; , &
	|g_{LR,NP}^{\mu\mu ee}|&<0.030~\text{\cite{Davidson:2003ha,Barranco:2007ej}}\; , \\\label{eq:NSI2}
	-0.16<g_{LL,NP}^{\tau\tau ee}&<0.11~\text{\cite{Bolanos:2008km}}\; , &
	-0.25<g_{LR,NP}^{\tau\tau ee}&<0.43~\text{\cite{Barranco:2007ej}}\;.
\end{align}
The search for $\bar\nu_\mu\to\bar\nu_e$ oscillations at zero distance in the KARMEN experiment~\cite{Eitel:2000by} can be recast in a constraint on~\cite{Biggio:2009nt}
\begin{align}\label{eq:NSI3}
	|g_{LL,NP}^{\mu e\mu e}|&<0.025\; , &
	|g_{LR,NP}^{\mu e\mu e}|&<0.025\;.
\end{align}
Numerically, the constraints are weak compared with constraints from lepton flavor universality and electroweak precision observables discussed above, but complementary.
Requiring the Wilson coefficients to stay within $2\sigma$ of the experimental errors, we translate the bounds in Eqs.~(\ref{eq:NSI1}), (\ref{eq:NSI2}) and (\ref{eq:NSI3}) to the constraints relevant for the comparison with the collider study of the bileptons in the third and fourth columns of  Table~\ref{tab:GF}. These constraints are not shown in the final Figs.~\ref{H} and \ref{D}, since they are weaker compared to constraints from the Fermi constant and lepton flavor universality.

\subsection{Existing collider constraints}
\label{sec:collcons}

The DELPHI collaboration interpreted their searches for $e^+e^-\to \ell^+\ell^-$ in terms of 4-lepton operators~\cite{Abdallah:2005ph} which are defined by the effective Lagrangian
\begin{equation}
\mathcal{L}_{eff} = \frac{g^2}{(1+\delta_{e\ell})\Lambda^2} \sum_{i,j=L,R} \eta_{ij} \bar e_i\gamma_\mu e_i\bar \ell_j \gamma^\mu \ell_j\;,
\end{equation}
where $\Lambda$ denotes the scale of the effective operator, $g$ is the coupling and $\eta_{ij}$ parameterizes which operators are considered at a given time and the relative sign of the operators in order to distinguish constructive (destructive) interference with the SM contribution. Conservative limits on the new physics scalars are obtained by setting the coupling to $g^2=4\pi$ and are summarized in the Table~30 of Ref.~\cite{Abdallah:2005ph}.

These constraints can be directly applied to bileptons by comparing the effective Lagrangians. The relevant 4-lepton operators are given by
\begin{align}
	\mathcal{L}_{eff} (H_{1,3}^{(\prime)})&=
	\frac{y_{1,3}^{ee}y_{1,3}^{\ell\ell}+y_{1,3}^{e\ell}y_{1,3}^{\ell e}}{(1+3\delta_{e\ell}) m_{H_{1,3}^0}^2} [\bar e \gamma_\mu P_{L} e] [\bar \ell \gamma^\mu P_{L}\ell]
	+\frac{y_{1}^{\prime ee}y_{1}^{\prime\ell\ell}+y_{1}^{\prime e\ell}y_{1}^{\prime\ell e}}{(1+3\delta_{e\ell}) m_{H_{1}^{\prime 0}}^2} [\bar e \gamma_\mu P_R e] [\bar \ell \gamma^\mu P_R\ell]\; ,
\\
\mathcal{L}_{eff}(H_2) &=
\frac{1}{2(1+3\delta_{e\ell})}
\left(\frac{1}{m_{h_2}^2}+\frac{1}{m_{a_2}^2}\right)
\left[
y_2^{ee}y_2^{\ell\ell*} [\bar e P_{R} e] [\bar \ell P_{L}\ell]
-\frac{|y_2^{e\ell}|^2}{2} [\bar e \gamma_\mu P_{L} e] [\bar \ell \gamma^\mu P_{R}\ell]
\right]
\nonumber\\&
+\frac{1}{4(1+3\delta_{e\ell})}
\left(\frac{1}{m_{h_2}^2}-\frac{1}{m_{a_2}^2}\right)
\Big[
	y_2^{ee}y_2^{\ell\ell} [\bar e P_R e] [\bar\ell P_R \ell]
\nonumber\\&
-\frac{y_2^{e\ell}y_2^{\ell e}}{2} \left([\bar e P_R e] [\bar\ell P_R \ell] +4 [\bar e \sigma^{\mu\nu}P_R e] [\bar\ell \sigma_{\mu\nu} P_R \ell]\right)
	+ (P_R\to P_L, y_2\to y_2^*)
\Big]
+[e\leftrightarrow\ell]
\nonumber\\ &\stackrel{m_{h_2}=m_{a_2}}{=}
\frac{1}{(1+3\delta_{e\ell})m_{h_2}^2}
\left[
y_2^{ee}y_2^{\ell\ell*} [\bar e P_{R} e] [\bar \ell P_{L}\ell]
-\frac{|y_2^{e\ell}|^2}{2} [\bar e \gamma_\mu P_{L} e] [\bar \ell \gamma^\mu P_{R}\ell]
+(e\leftrightarrow \ell)\right]\; ,
\label{H2LEP}
\\
\mathcal{L}_{eff}(\Delta_{1,3}) &= \frac{2|\lambda_{1,3}^{e\ell}|^2}{(1+3\delta_{e\ell})m_{\Delta_{1,3}^{++}}^2} [\bar e \gamma_\mu P_{R,L} e] [\bar \ell \gamma^\mu P_{R,L} \ell]\; .
\end{align}
As we have demonstrated in Ref.~\cite{Li:2018cod}, the analysis of contact interactions in Ref.~\cite{Abdallah:2005ph} does not directly apply to $\Delta_{2\mu}^{++}$, because the induced effective interactions do not fall into any of the types of effective interactions considered in Ref.~\cite{Abdallah:2005ph}. Similarly for $H_2$, the analysis only applies in the limit of degenerate neutral (pseudo)scalar masses ($m_{h_2}=m_{a_2}$) and in the absence of one of the diagonal entries $y_2^{ee,\ell\ell}$, such that the scalar operator in the first term of Eq.~(\ref{H2LEP}) is not induced.
	
For the other operators we list the translated limits for masses well above the center-of-mass energy of LEP, $\sqrt{s}\sim 130-207$ GeV, in Table~\ref{tab:LEPbounds}. Note that these limits are only valid when the new particle mass is much greater than $\sqrt{s}$.
To make it valid for any masses, we should replace the mass in Table~\ref{tab:LEPbounds} by $(s\cos\theta/2+m^2)^{1/2}$ after averaging over the scattering angle $\langle\cos\theta\rangle\simeq 1/2$.

\begin{table}[tb]\centering
\resizebox{\linewidth}{!}{\begin{tabular}{|c|c|c|c|}\hline
 & $e^+e^- \to e^+ e^-$ & $e^+e^- \to \mu^+\mu^-$ & $e^+e^- \to \tau^+\tau^-$\\\hline
	$H_{1,3}$ & $|y_{1,3}^{ee}|\leq 6.7\times 10^{-4} m_{H_{1,3}^0}$ & $\sqrt{|y_{1,3}^{ee}y_{1,3}^{\mu\mu} + y_{1,3}^{e\mu}y_{1,3}^{\mu e}|}\leq 4.9\times 10^{-4} m_{H_{1,3}^0}$ & $\sqrt{|y_{1,3}^{ee}y_{1,3}^{\tau\tau}+y_{1,3}^{e\tau}y_{1,3}^{\tau e}|}\leq 4.5\times 10^{-4} m_{H_{1,3}^0}$
\\\hline
$H_1^{\prime0}$ & $|y_1^{\prime ee}|\leq 6.8\times 10^{-4} m_{H_1^{\prime 0}}$ & $\sqrt{|y_1^{\prime ee}y_1^{\prime\mu\mu}+y_1^{\prime e\mu}y_1^{\prime\mu e}|}\leq 5.1\times 10^{-4} m_{H_1^{\prime 0}}$& $\sqrt{|y_1^{\prime ee}y_1^{\prime \tau\tau}+y_1^{\prime e\tau}y_1^{\prime\tau e}|}\leq 4.7\times 10^{-4} m_{H_1^{\prime 0}}$
\\\hline
$H_2$ & $|y_2^{ee}|\leq 5.3\times 10^{-4} m_{h_2}$ & $|y_2^{e\mu}|\leq 2.5\times 10^{-3} m_{h_2}$& $|y_2^{e\tau}|\leq 2.4\times 10^{-3} m_{h_2}$
\\\hline
$\Delta_1^{++}$ & $|\lambda_1^{ee}|\leq 6.8\times 10^{-4} m_{\Delta_1^{++}}$ & $|\lambda_1^{e\mu}|\leq 3.6\times 10^{-4} m_{\Delta_1^{++}}$ & $|\lambda_1^{e\tau}|\leq 3.3\times 10^{-4} m_{\Delta_1^{++}}$
\\\hline
$\Delta_3$ & $|\lambda_3^{ee}|\leq 6.7\times 10^{-4} m_{\Delta_3^{++}}$ & $|\lambda_3^{e\mu}|\leq 3.4\times 10^{-4} m_{\Delta_3^{++}}$ & $|\lambda_3^{e\tau}|\leq 3.2\times 10^{-4} m_{\Delta_3^{++}}$
\\\hline
\end{tabular}}
\caption{LEP limits on couplings for masses well above the center of mass energy $\sqrt{s}\sim 130-207$ GeV. The limits for $H_2$ from $e^+e^-\to\mu^+\mu^-, \tau^+\tau^-$, i.e. $e\neq \ell$, are obtained under the assumption that $y_2^{ee}$ or $y_{2}^{\ell\ell}$ vanishes. }
\label{tab:LEPbounds}
\end{table}

Most available searches for a singly-charged scalar as well as a second neutral heavy Higgs at the LHC generally do not apply here, because they rely on couplings to quarks. If the singly-charged scalar is part of a scalar multiplet where the neutral component obtains a vacuum expectation value, the analyses in Refs.~\cite{Aad:2015nfa,Sirunyan:2017sbn} place a constraint on the (electroweak) vector boson fusion production cross section and the subsequent decay to electroweak gauge bosons for singly-charged scalars with masses in the range $200-2000$ GeV.
Both neutral and singly-charged scalars may also be produced in pairs via electroweak processes, but there are no applicable general searches to our knowledge. They may also be produced via $s$-channel $W$ boson exchange together with another component in the electroweak multiplet.

There are searches for doubly-charged scalars produced via electroweak pair production in both ATLAS and CMS experiments.
The most stringent limits for decays to $e^\pm e^\pm$, $\mu^\pm \mu^\pm$, $e^\pm \mu^\pm$ pairs are set by the ATLAS experiment~\cite{Aaboud:2017qph}. It excludes masses $m_{\Delta^{++}_{1(3)}}\leq 320 (450)$ GeV assuming BR$(\Delta^{++}_{1,3} \to \ell^+ \ell^+)\geq 10\%$. Assuming 100\% branching ratio for a given channel, the constraints range from $650$ GeV for $\Delta_1^{++}\to e^+ e^+$ to $850$ GeV for $\Delta_3^{++}\to \mu^+\mu^+$. Although the constraints set by the CMS experiment are slightly lower, it sets the most stringent lower limits on the final states with $\tau$ leptons~\cite{CMS:2017pet}. Assuming 100\% branching ratio in each channel, they range from $535$ GeV for $\Delta_3^{++}\to \tau^+\tau^+$ to $714$ GeV for $\Delta_3^{++}\to \tau^+ e^+$.

\section{Sensitivity of future experiments to CLFV}
\label{sec:sen}

\subsection{Sensitivity from neutrino trident production}
Neutrino trident production, the production of a charged lepton pair from a neutrino scattering off the Coulomb field of a nucleus, provides an interesting signature to search for new physics beyond the SM~\cite{Mishra:1991bv,Gaidaenko:2000hg,Altmannshofer:2014pba}.
So far, only the muonic trident has been measured with the results of $\sigma_{\rm exp}/\sigma_{\rm SM}=1.58\pm0.64$ at CHARM-II~\cite{Geiregat:1990gz}, $\sigma_{\rm exp}/\sigma_{\rm SM}=0.82\pm0.28$ at CCFR~\cite{Mishra:1991bv} and $\sigma_{\rm exp}/\sigma_{\rm SM}=0.72^{+1.73}_{-0.72}$ at NuTeV~\cite{Adams:1999mn}. While CHARM-II and CCFR achieved an accuracy of the level of 35\%~\cite{Altmannshofer:2019zhy}, their measurements agree with the SM prediction and no signal has been established at NuTeV.

This will be improved by a measurement at the near detector of the Deep
Underground Neutrino Experiment (DUNE), which can reach an accuracy of
25\%~\cite{Altmannshofer:2019zhy}. See also Ref.~\cite{Ballett:2018uuc} for a
related study. The DUNE near detector is expected to measure three neutrino
trident channels: $\nu_\mu N  \to \nu_\mu e^+ e^- N$, $\nu_\mu N \to \nu_\mu
\mu^+ \mu^- N$ and $\nu_\mu N\to \nu_e e^+ \mu^- N$. The third one is not
sensitive to new physics in a scheme where the Fermi constant $G_F$ is
determined by muon decay, as it is directly related to muon decay by crossing
symmetry. We calculate the cross sections of the former two channels in
presence of the new contributions to the effective operators in
Eq.~\eqref{eq:NSI}, using the code provided by
Ref.~\cite{Altmannshofer:2019zhy}. Assuming a precision of 25\% for the cross
section measurements, one can translate the expectations of the Wilson
coefficients in Eqs.~(\ref{gLL}) and (\ref{gLR}) into the sensitivities to the
CLFV couplings quoted in Table~\ref{tab:trident}. Note that all new physics
contributions to the trident process $\nu_\mu N \to \nu_\mu e^+ e^- N$ in
principle result in
two disconnected allowed regions of parameter space if no
signal is observed at DUNE. However, some of them are not accessible by
interactions of the bileptons and only one of the two regions is theoretically
reasonable. We find the two reasonable regions for $H_1^0$ and $\Delta_2$ as shown
in the left column of Table~\ref{tab:trident}.

\begin{table}[tb]
	\begin{tabular}{|c|c|c|}\hline
		& $\nu_\mu N \to \nu_\mu e^+e^- N$
		& $\nu_\mu N\to \nu_\mu \mu^+\mu^- N$
		\\\hline
		\multirow{2}{*}{$H_1^0$}
		& $-2.0\times 10^{-5} m_{H_1^0}^2 \leq y_1^{\mu\mu}y_1^{ee}\leq -1.6\times 10^{-5} m_{H_1^0}^2$
		& \multirow{2}{*}{$
	(y_1^{\mu\mu})^2\leq 3.4\times 10^{-6} m_{H_1^0}^2$}
		\\
		& $-2.2\times 10^{-6} m_{H_1^0}^2 \leq y_1^{\mu\mu}y_1^{ee}\leq 1.8\times 10^{-6} m_{H_1^0}^2$
		&
		\\\hline
	$H_2$
&$|y_2^{e\mu}|^2\leq 3.9 \times 10^{-6} m_{H_2^+}^2$
	&$|y_2^{\mu\mu}|^2\leq 2.0 \times 10^{-5} m_{H_2^+}^2$
		\\\hline
		$H_3$
		&
		$(y_3^{e\mu})^2\leq 8.8\times 10^{-7} m_{H_3^+}^2$
		&
	$(y_3^{\mu\mu})^2\leq 1.7\times 10^{-6} m_{H_3^+}^2$
		\\\hline
		\multirow{2}{*}{$\Delta_2$}
& $|\lambda_2^{e\mu}|^2\leq 2.7\times 10^{-6} m_{\Delta_2^+}^2$
& \multirow{2}{*}{$|\lambda_2^{\mu\mu}|^2\leq 1.4\times 10^{-5} m_{\Delta_2^+}^2$}
		\\
& $1.3\times 10^{-5} m_{\Delta_2^+}^2 \leq |\lambda_2^{e\mu}|^2\leq 1.7\times 10^{-5} m_{\Delta_2^+}^2$
&
		\\\hline
	$\Delta_3$
	&
		$|\lambda_3^{e\mu}|^2\leq 1.8\times 10^{-6} m_{\Delta_3^+}^2$
		&
	$|\lambda_3^{\mu\mu}|^2\leq 3.4\times 10^{-6} m_{\Delta_3^+}^2$
		\\\hline
	\end{tabular}
\caption{Sensitivity reach from neutrino trident production in units of GeV$^{-2}$ assuming 25\% precision for the measurement of the cross section. In the absence of any deviation from the SM neutrino trident cross section, the Yukawa couplings of the bileptons have to satisfy the above-listed constraints.
}
\label{tab:trident}
\end{table}

\subsection{Sensitivity of future lepton colliders to the CLFV}
Apart from studying rare decays, the CLFV can be probed through scattering processes at lepton colliders. The new particles beyond the SM either mediate the scattering in off-shell channels or can be produced on-shell.
In this work we focus on on-shell production of the bileptons together with a pair of different flavor leptons. The benefit of this on-shell scenario is that it only depends on one single CLFV coupling in each production channel and can be directly compared with the constraints placed by the low-energy experiments.

The proposed lepton colliders, in terms of the center of mass (c.m.) energy and the integrated luminosity used in our analysis, are
\begin{itemize}
\item Circular Electron Positron Collider (CEPC): 5 ab$^{-1}$ at 240 GeV~\cite{CEPCStudyGroup:2018ghi},
\item Future Circular Collider (FCC)-ee: 16 ab$^{-1}$ at 240 GeV~\cite{FCCee:2017},
\item International Linear Collider (ILC): 4 ab$^{-1}$ at 500 GeV~\cite{Barklow:2015tja}, 1 ab$^{-1}$ at 1 TeV~\cite{Baer:2013cma},
\item Compact Linear Collider (CLIC): 5 ab$^{-1}$ at 3 TeV~\cite{Charles:2018vfv}.
\end{itemize}
The CLFV processes can happen through the scattering of $e^+
e^-$~\cite{Dev:2017ftk,Sui:2017qra,Dev:2018upe} with on-shell new particles in
final states, i.e. $e^+ e^-\to \ell^\pm_i\ell^\mp_j H^0, \ell^\pm_i\ell^\pm_j
\Delta^{\mp\mp}$ with $\ell_i,\ell_j=e,\mu,\tau$. The CLFV channels via $\Delta
L=0$ or $\Delta L=2$ interaction at an $e^+e^-$ collider for probing the couplings
$y^{ij}, \lambda^{ij}$ are given in Table~\ref{tab:opposite-sign}. The
processes with one electron or position in final states occur through both s
and t channels mediated by $Z/\gamma^\ast$. The processes without $e^\pm$ in
final states only occur in s channel.

\begin{table}[tb!]
\begin{center}
\begin{tabular}{|c|c|c|}
        \hline
        flavor $ij$ & $\Delta L=0$ CLFV channel & $\Delta L=2$ CLFV channel \\
        \hline
        $e\mu$ & $e^+e^-\to e^\pm\mu^\mp H^0$ (s+t) & $e^+e^-\to e^\pm\mu^\pm \Delta^{\mp\mp}$ (s+t) \\
        \hline
        $e\tau$ & $e^+e^-\to e^\pm\tau^\mp H^0$ (s+t) & $e^+e^-\to e^\pm\tau^\pm \Delta^{\mp\mp}$ (s+t) \\
        \hline
        $\mu\tau$ & $e^+e^-\to \mu^\pm\tau^\mp H^0$ (s) & $e^+e^-\to \mu^\pm\tau^\pm \Delta^{\mp\mp}$ (s) \\
        \hline
\end{tabular}
\end{center}
\caption{CLFV channels via $\Delta L=0$ or $\Delta L=2$ interaction at $e^+e^-$ collider, for probing coupling $y^{ij}, \lambda^{ij}$.
}
\label{tab:opposite-sign}
\end{table}

In order to estimate the lepton collider sensitivity to the CLFV couplings, we
create UFO model files using \texttt{FeynRules}~\cite{Alloul:2013bka} and interface them
with \texttt{MadGraph5\_aMC@NLO}~\cite{Alwall:2014hca} to generate signal events. We
apply basic cuts $p_T>10$ GeV and $|\eta|<2.5$ on the leptons in final states and assume a tau efficiency of $60\%$~\cite{Baer:2013cma}. Thus, the sensitivity reach is weakened by a factor 1.3 for the channels with one tau lepton in the final state compared with the reach for the $e\mu$ channel.
The CLFV processes are not triggered by the initial state radiation (ISR) or final state radiation (FSR), and ISR/FSR barely introduces significant systematic uncertainties for our LFV signal. Moreover, the observation of CLFV does not rely on the exhibition of high energy tail induced by the radiation effects (ISR, FSR, bremsstrahlung, beamstrahlung, etc.). We thus neglect ISR/FSR effects in our analysis.

Besides the charged lepton pairs, the new bosons can decay into other SM particles, which makes the reconstruction rather model-dependent.
To give a model-independent prediction, we assume 10\% efficiency for the reconstruction of the new bosons. This takes into account the
effects of cuts on decay products of bileptons, their decay branching fraction, and possibly missing bileptons in the detector.
The dominant SM background is the Higgsstrahlung process $e^+e^-\to Zh$ followed by the mis-identification of one charged lepton from $Z$ boson decay~\cite{Dev:2017ftk}. The invariant mass of the two charged leptons in our signal can be easily distinguished from the background with a $Z$ resonance peak. Thus, after vetoing the $Z$ mass window, our CLFV signal is almost background free.
We take the significance of $S/\sqrt{S+B}\approx \sqrt{S}$ as 3 for the observation of CLFV.

\begin{figure}[tbp!]
\begin{center}
\includegraphics[scale=1,width=5.4cm]{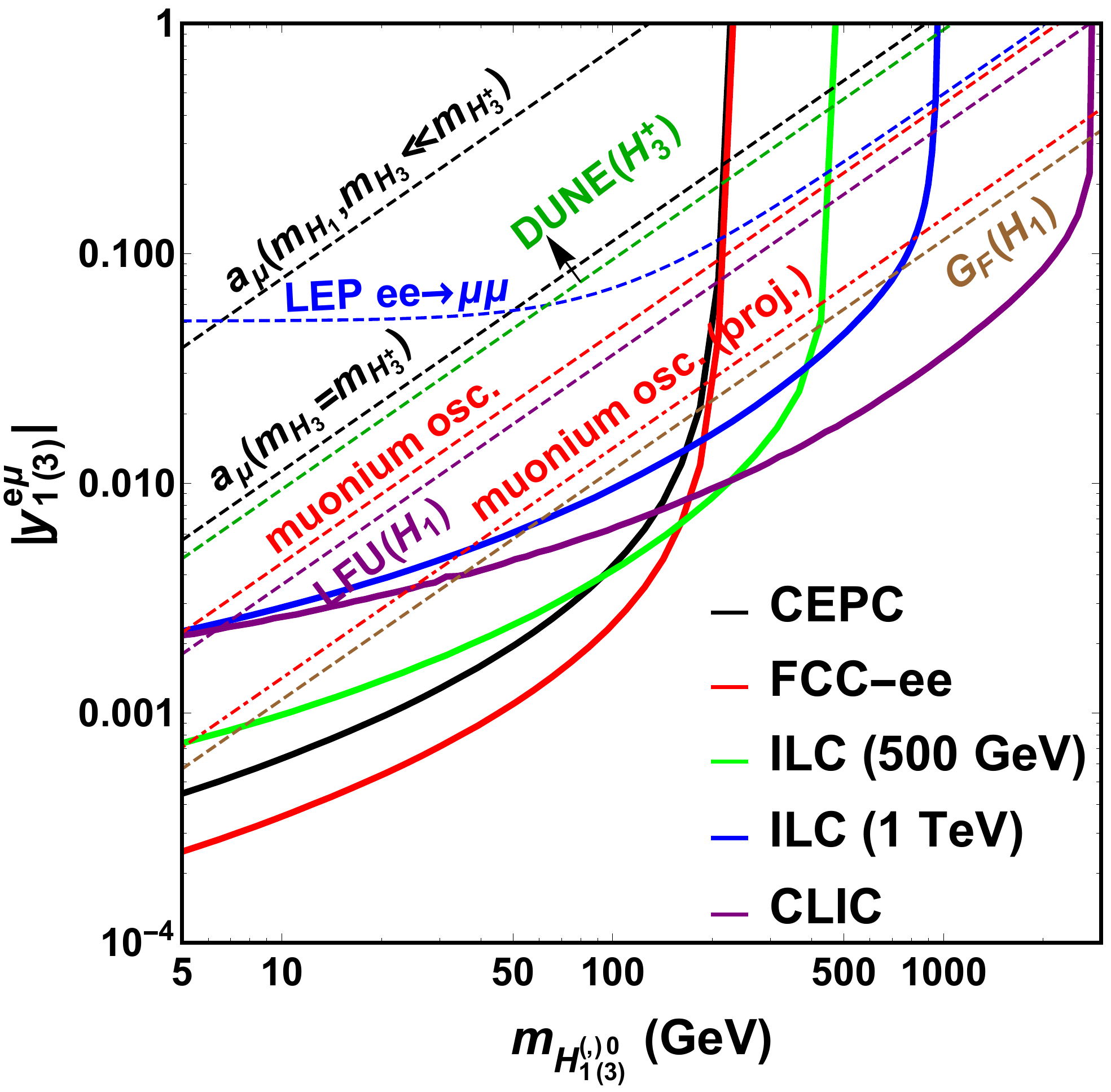}
\includegraphics[scale=1,width=5.4cm]{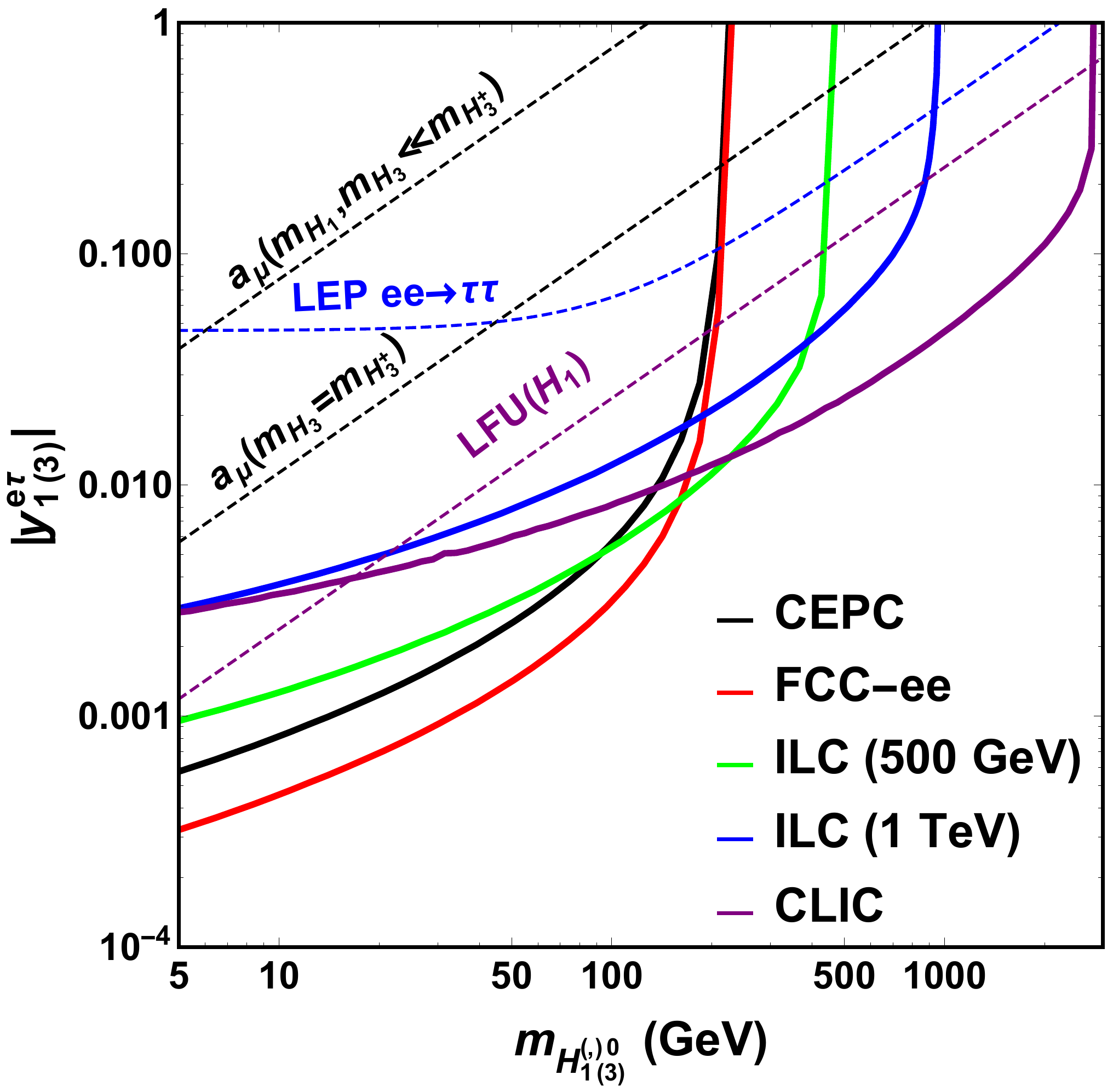}
\includegraphics[scale=1,width=5.4cm]{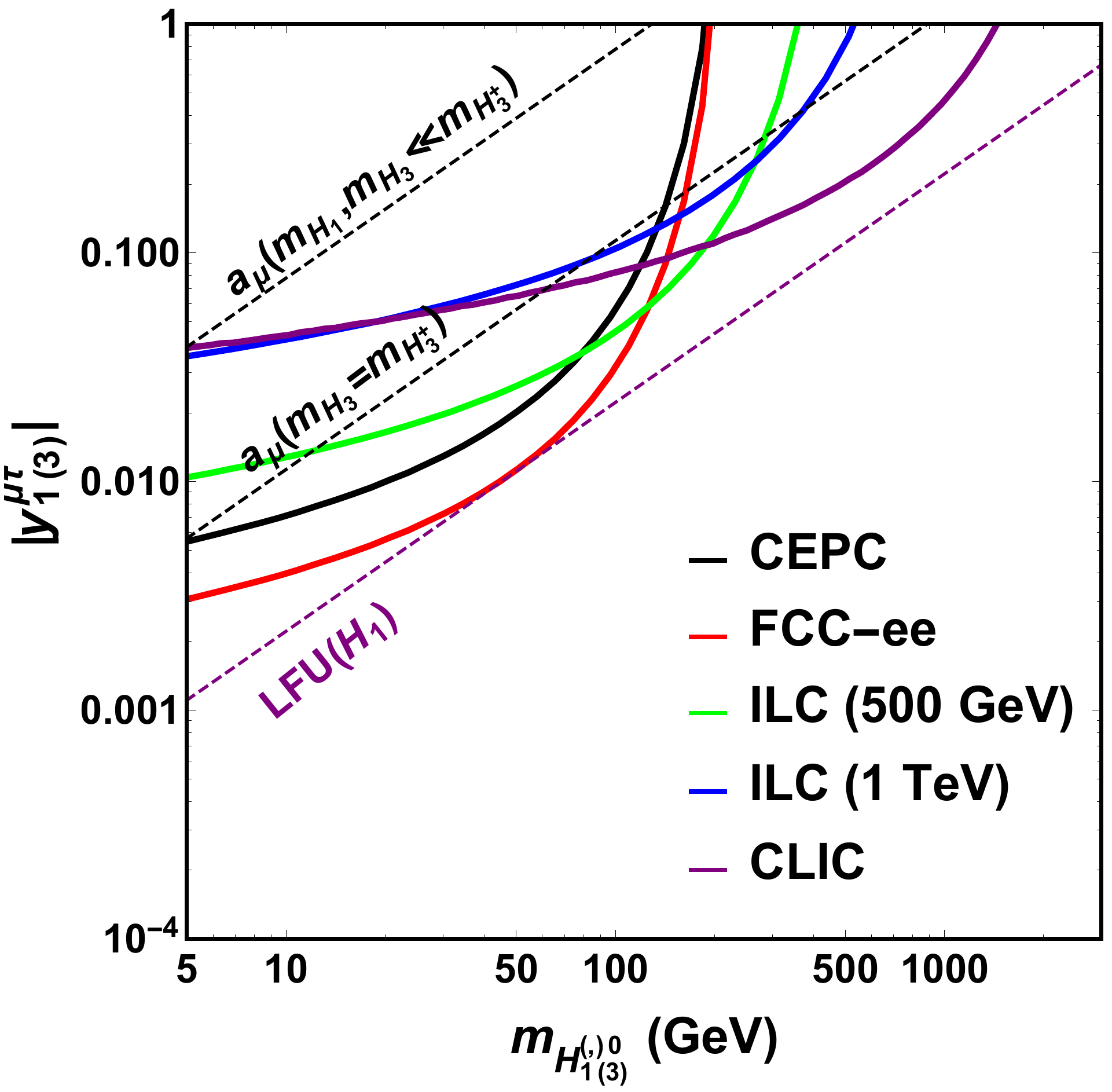}\\
\includegraphics[scale=1,width=5.4cm]{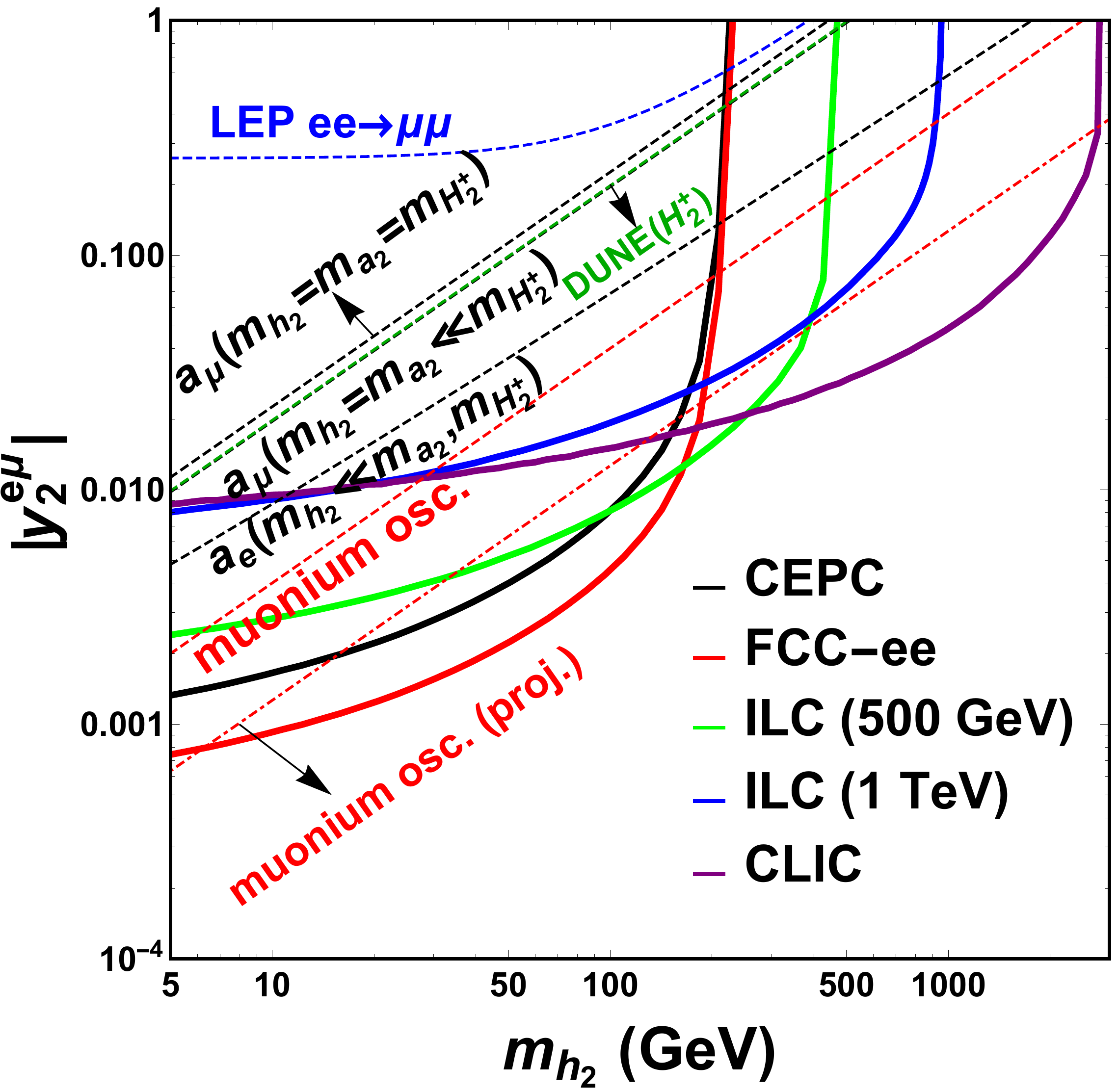}
\includegraphics[scale=1,width=5.4cm]{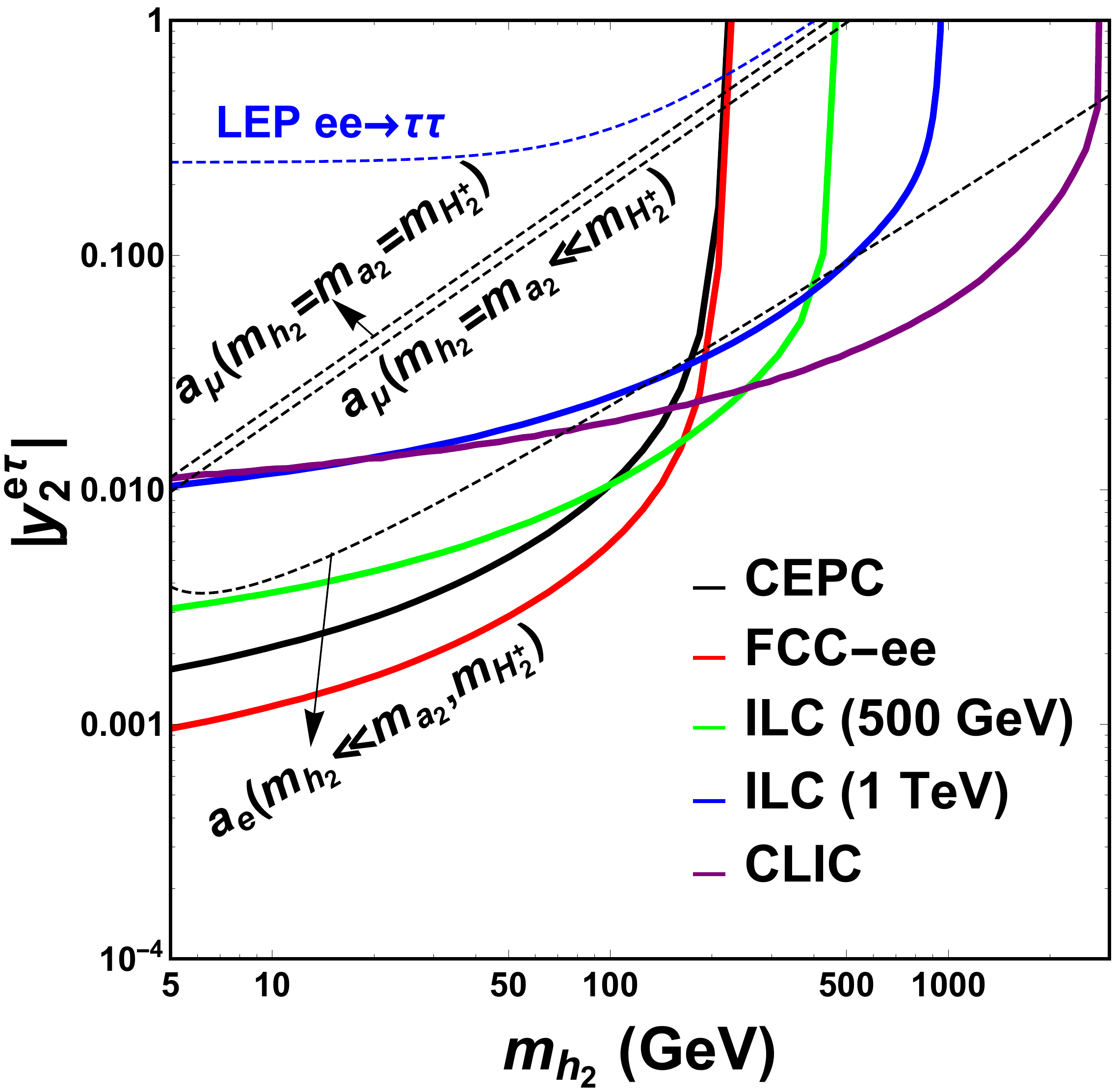}
\includegraphics[scale=1,width=5.4cm]{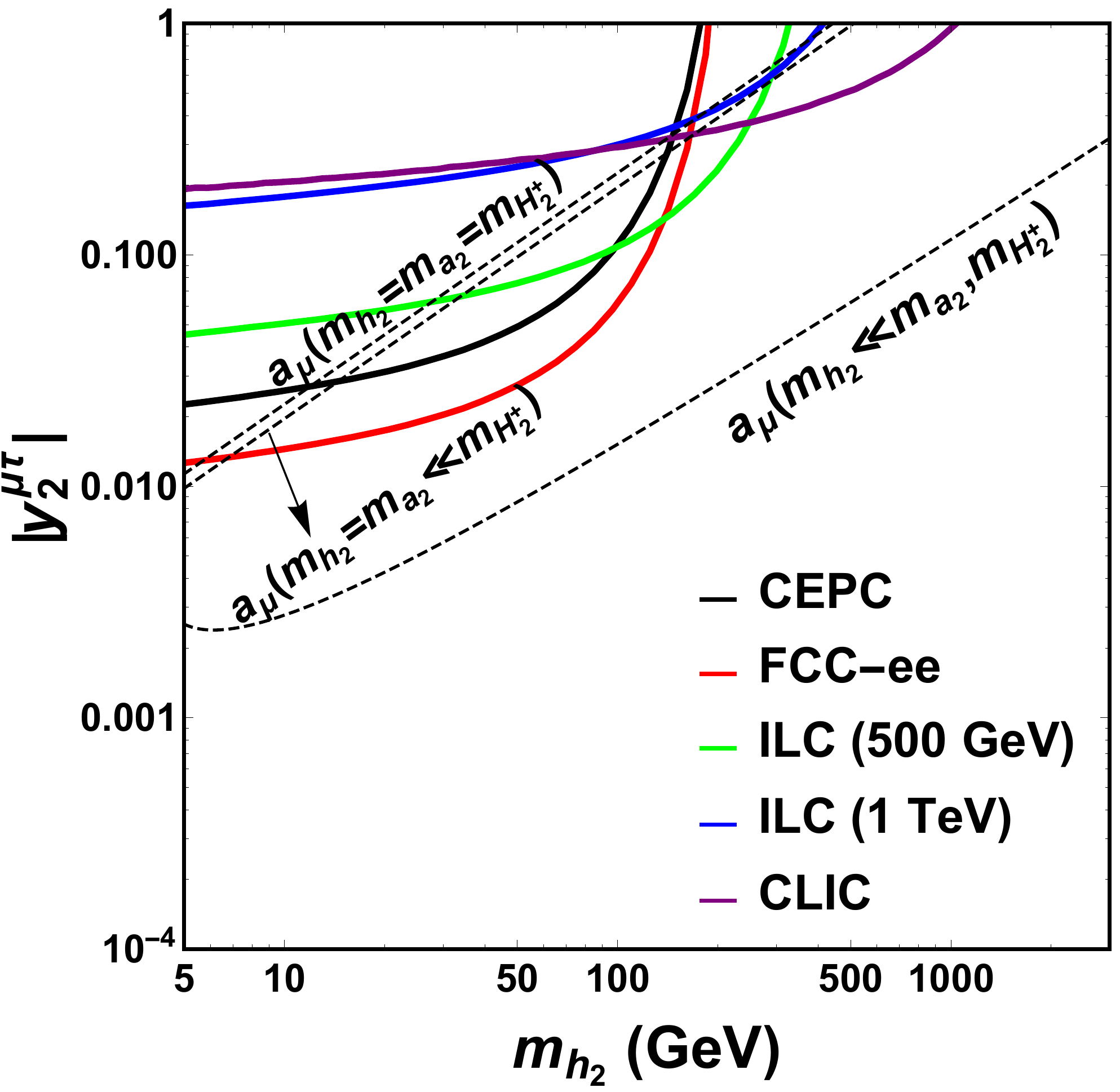}
\end{center}
\caption{
	Sensitivity to $|y^{e\mu}|$ (left), $|y^{e\tau}|$ (middle) coupling through $e^+e^-\to e^\pm\mu^\mp (e^\pm\tau^\mp)+H^0$, and $|y^{\mu\tau}|$ (right) coupling through $e^+e^-\to \mu^\pm\tau^\mp+H^0$, for $H_{1}^{(\prime)0}, H_3^0$ (top) and $H_2^0$ (bottom) interactions. For the $H_2$ case, we assume either $h_2$ or $a_2$ is produced. The bounds from low-energy experiments are shown as dashed lines. The projected sensitivity reach from a future muonium-antimuonium conversion experiment is shown as a dot-dashed line. The green dashed line indicates the sensitivity reach of neutrino trident production at the DUNE near detector.
}
\label{H}
\end{figure}

\begin{figure}[tbp!]
\begin{center}
\includegraphics[scale=1,width=5.4cm]{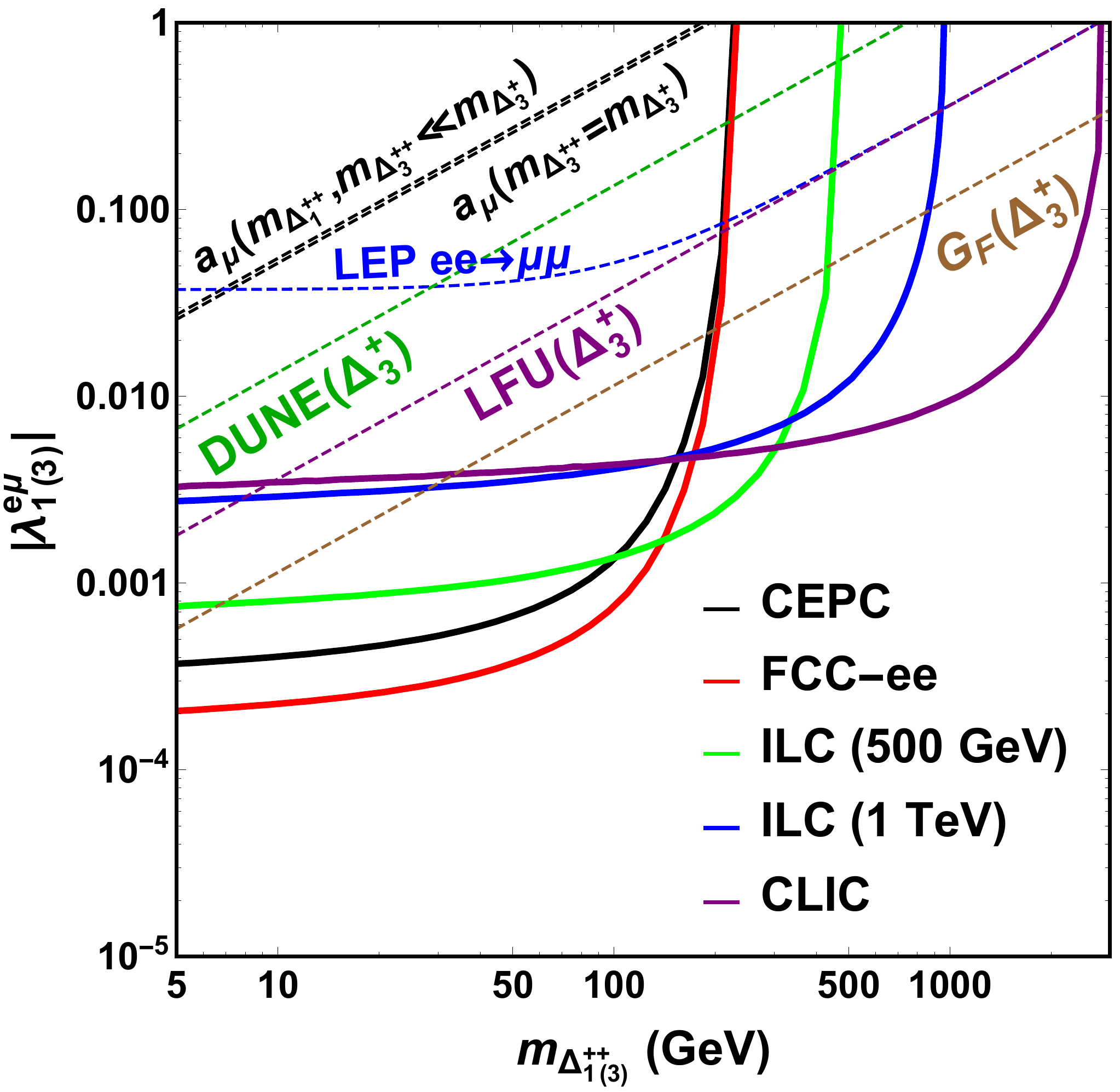}
\includegraphics[scale=1,width=5.4cm]{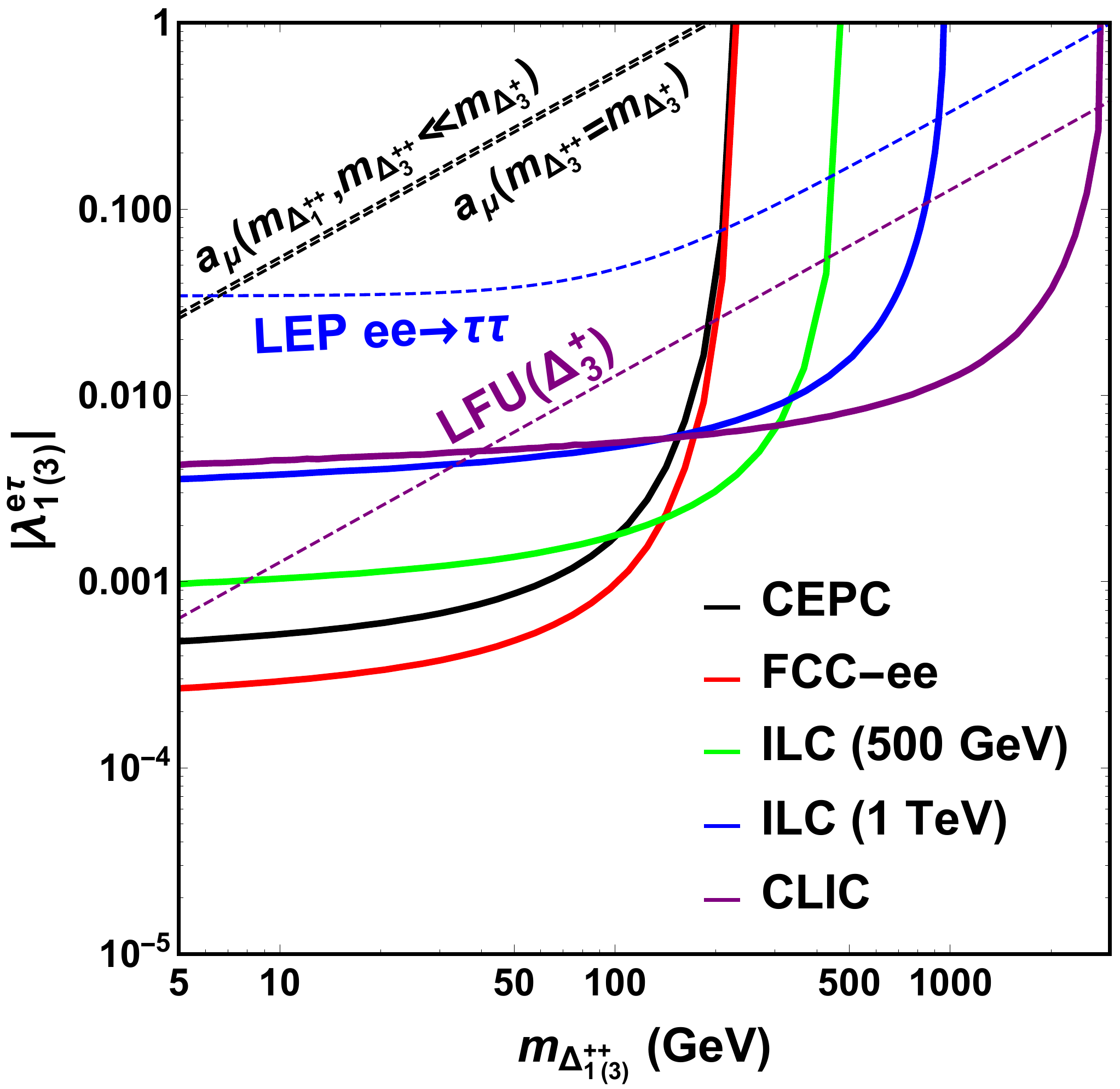}
\includegraphics[scale=1,width=5.4cm]{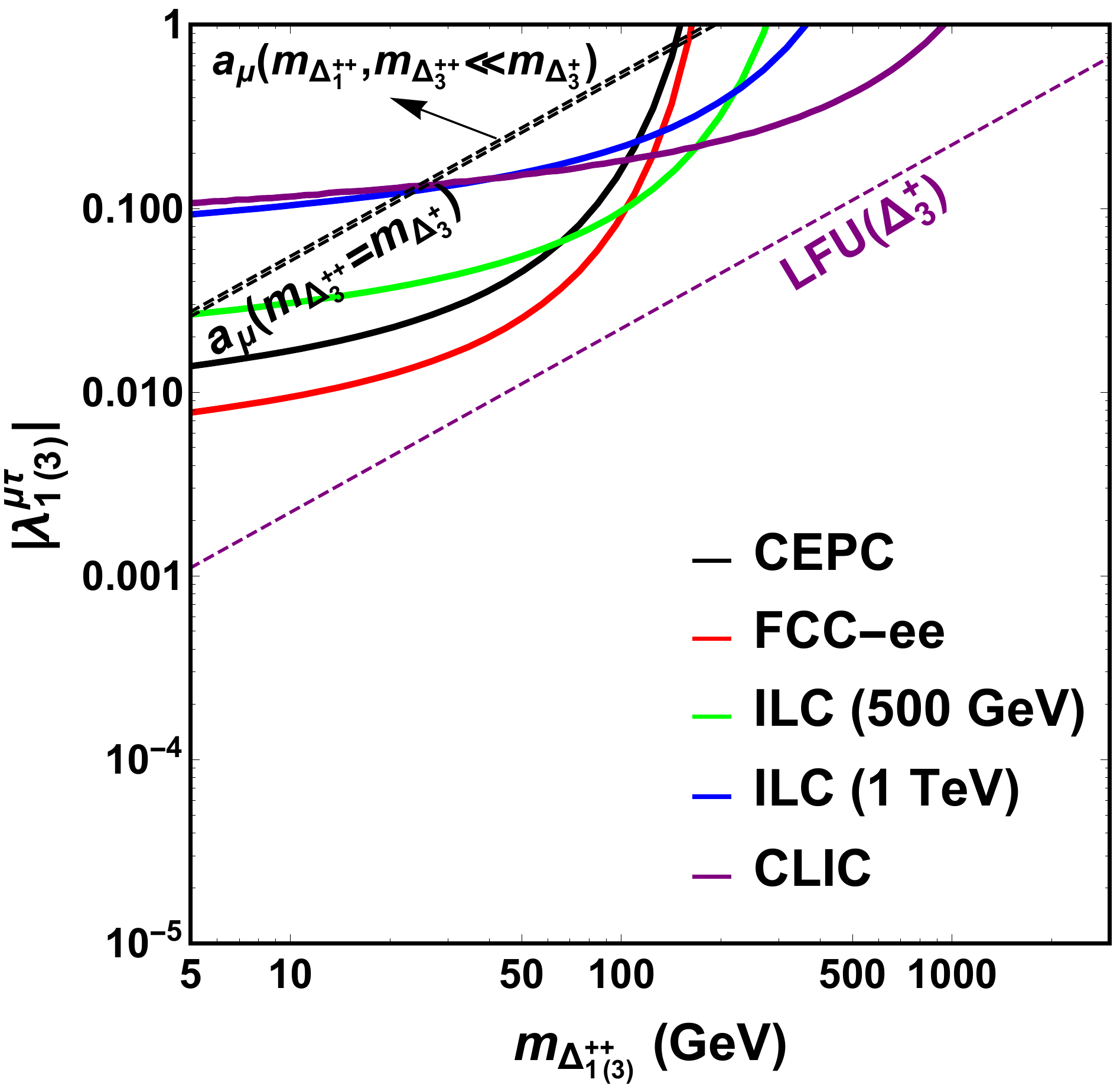}\\
\includegraphics[scale=1,width=5.4cm]{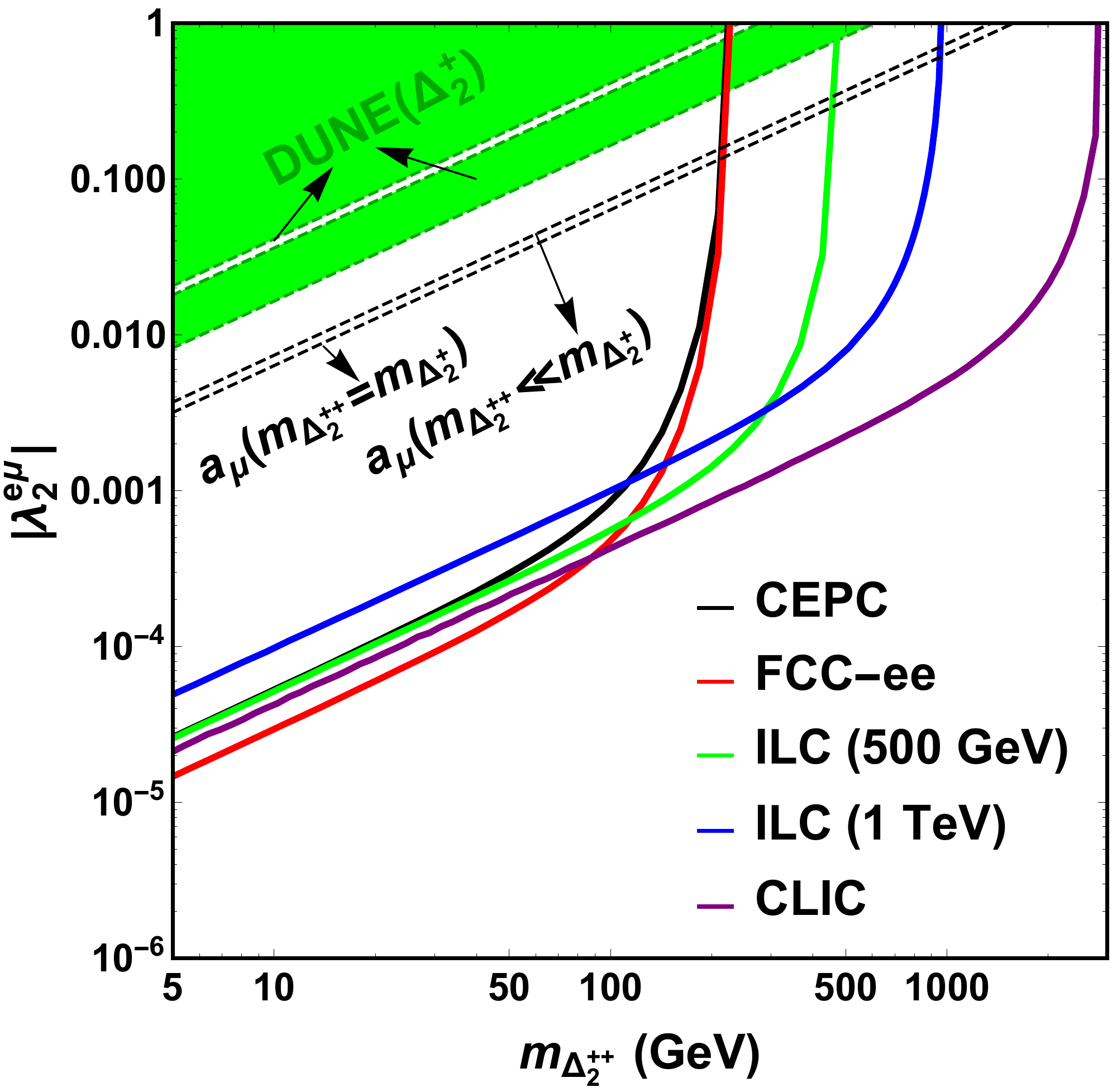}
\includegraphics[scale=1,width=5.4cm]{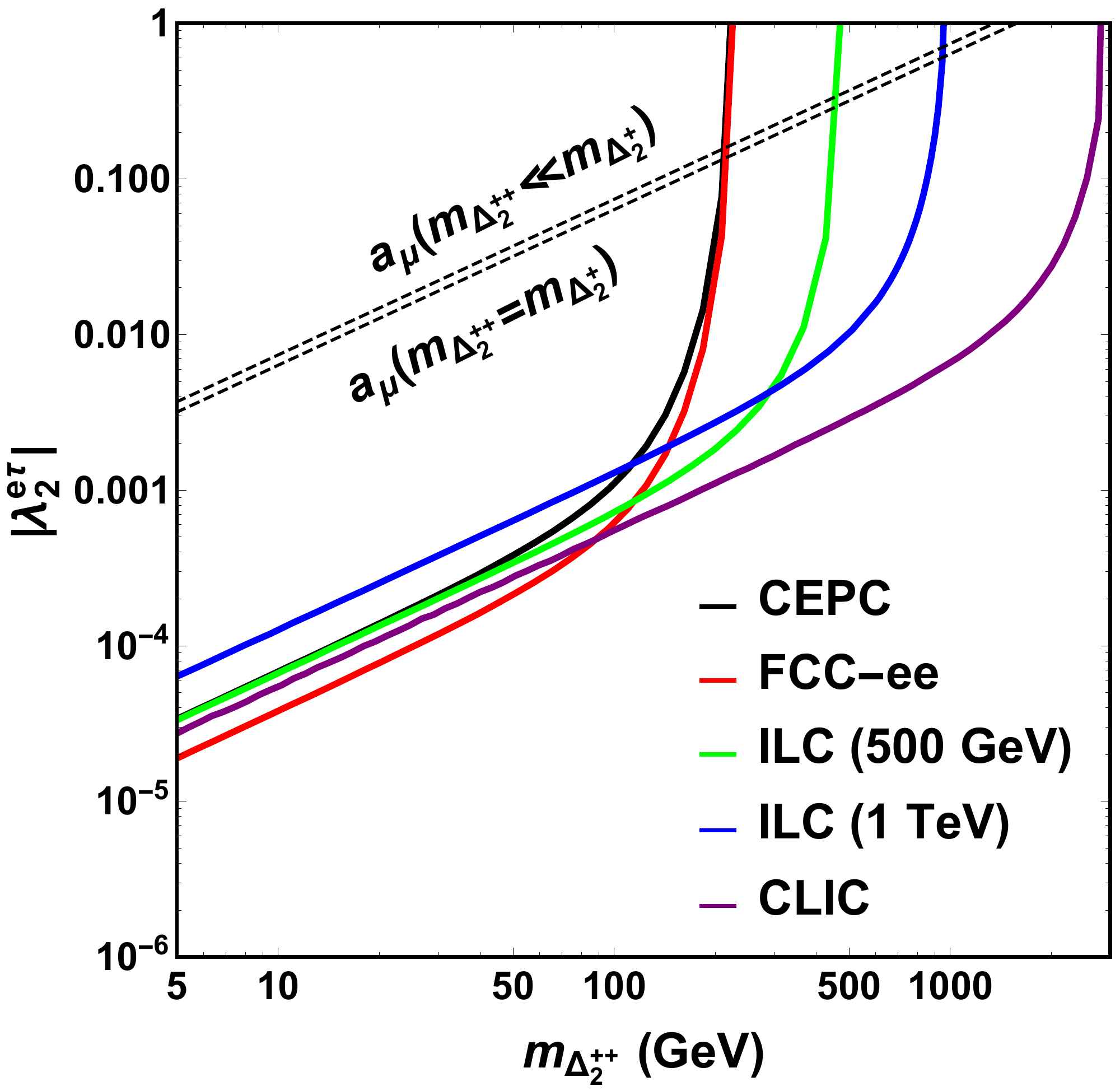}
\includegraphics[scale=1,width=5.4cm]{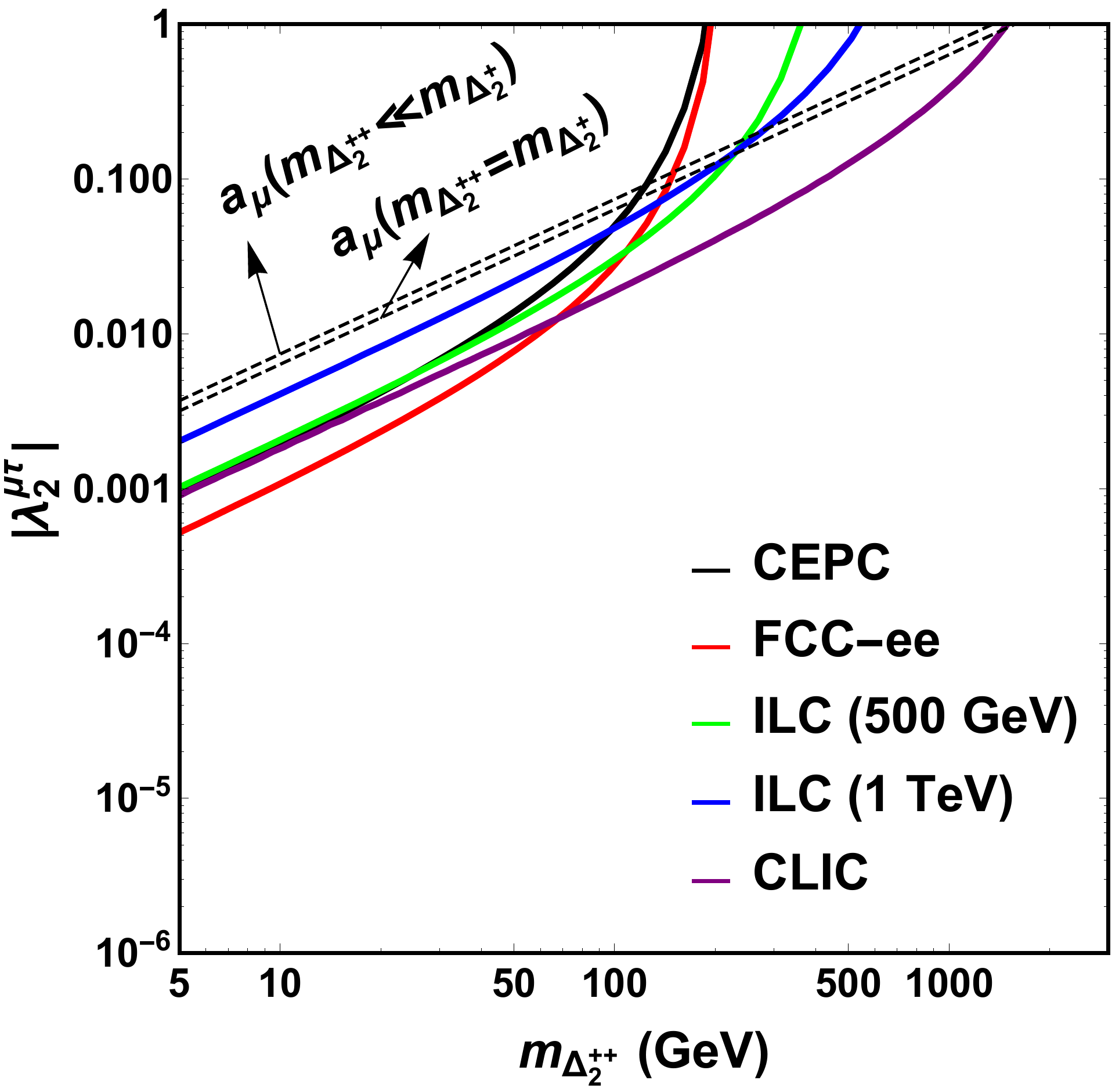}
\end{center}
\caption{
Sensitivity to $|\lambda^{e\mu}|$ (left), $|\lambda^{e\tau}|$ (middle) coupling through $e^+e^-\to e^\pm\mu^\pm (e^\pm\tau^\pm)+\Delta^{\mp\mp}$, and $|\lambda^{\mu\tau}|$ (right) coupling through $e^+e^-\to \mu^\pm\tau^\pm+\Delta^{\mp\mp}$, for $\Delta_{1,3}^{++}$ (top) and $\Delta_2^{++}$ (bottom) interactions. The bounds from low-energy experiments are shown as dashed lines. The green dashed line indicates the sensitivity reach of neutrino trident production at the DUNE near detector. For the $\lambda_2^{e\mu}$ coupling of $\Delta_2^+$, in the bottom-left panel, the sensitivity reach of neutrino trident production at the DUNE experiment is shaded in green to clearly indicate that DUNE is not sensitive to a narrow range of parameters between the green-shaded regions.
}
\label{D}
\end{figure}

We show the sensitivity to $\Delta L=0$ and $\Delta L=2$ couplings in
Figs.~\ref{H} and \ref{D}, respectively. Each of the CLFV processes only depends on one LFV coupling as shown in Table~\ref{tab:opposite-sign}. Thus, the plots Figs.~\ref{H} and \ref{D} do not rely on the values of other couplings. Note that in this work we do not
expect to distinguish the chiral nature of the couplings of the mediating
particles. Thus, the following results for vector $H_{1,3}^0$ and scalar
$\Delta_3^{++}$ which only couple to left-handed leptons are the same as those
for $H_1^{\prime 0}$ and $\Delta_1^{++}$ with only couplings to right-handed
leptons, respectively.
One can see that the interference between both s and t channels makes it more sensitive to probe couplings with $e\mu$ and $e\tau$ flavors, as shown in the left and middle panels of the figures. Smaller couplings can be reached for vector particles, such as $H_{1,3}^{(\prime)0}$ and $\Delta_2^{++}$, compared with new scalar bosons in both $\Delta L=0$ and $\Delta L=2$ interactions. The FCC-ee with the highest integrated luminosity provides the most sensitive environment and CEPC is the second most sensitive one in the low mass region. The CLIC and ILC with larger c.m.~energy can probe the high mass region of the new particles.

The relevant upper bounds or projected sensitivities from low-energy experiments are also displayed for the corresponding couplings. We do not show the constraints from LFU in $\tau$ decays and electroweak precision observables, if they are weaker than the ones from the AMMs. Unless stated otherwise we assume that the components of the bilepton multiplets are degenerate in mass. It is straightforward to generalize the constraints from the effective operators of two charged leptons and two neutrinos in Eq.~\eqref{eq:NSI} by shifting the contours horizontally depending on the mass splitting between the different components of the bilepton multiplets. The constraints from the anomalous magnetic moments require a reevaluation. Lepton flavor universality sets the most stringent
bound for $H_1^0$ and $\Delta_3$. The LFU bound excludes a majority of parameter space
that the future lepton colliders can reach for $y_1^{\mu\tau}$ and
$\lambda_3^{\mu\tau}$ couplings. Electroweak precision observables provide an even better constraint on the $e-\mu$ couplings of $H_1^0$ and $\Delta_3$. Muonium-antimuonium conversion also
provides strong constraint on the $y^{e\mu}$ couplings. The constraints from the lepton AMMs
vary with different mass spectra of the new particles and is relatively weak
unless there is only one visible neutral scalar in the $H_2$ case. Finally, the LEP constraints from $e^+ e^-\to \ell^+ \ell^-$ scattering are generally weaker than the constraints from low-energy precision experiments. The neutrino trident cross section measurement at the DUNE near detector is not expected to be able to probe new parameter space.

\subsection{Sensitivity of an improved muonium-antimuonium conversion experiment}

A future dedicated muonium-antimuonium conversion experiment may be able to improve the sensitivity to the Wilson coefficient of the effective operator by one order of magnitude~\cite{Bernstein:2019PC}. This directly translates to an improvement in sensitivity by one order of magnitude compared with the constraints listed in Table~\ref{tab:muonium} or about a factor of $3$ in terms of the CLFV couplings as shown in Fig.~\ref{H}. Note, although muonium-antimuonium conversion can not probe the CLFV couplings of $\Delta L=2$ bileptons, it is sensitive to combinations of flavor-diagonal couplings.

\section{Conclusion}
\label{sec:Con}

We studied the sensitivity of on-shell production of a bilepton with charged-lepton-flavor-violating couplings at future lepton colliders and compared it with the current constraints and future sensitivities of other experiments. We consider all possible scalar and vector bileptons with non-zero off-diagonal CLFV couplings. The bileptons are categorized into lepton number conserving ($\Delta L=0$) bileptons with $y^{ij}$ couplings and $\Delta L=2$ bileptons with $\lambda^{ij}$ couplings.

Depending on the nature of different bileptons, the most stringent constraints on the flavor off-diagonal couplings are from different measurements: Muonium-antimuonium conversions are currently most sensitive to the $e-\mu$ coupling of $\Delta L=0$ bileptons with the exception of $H_1^0$. Electroweak precision observables provide the most stringent constraint on the $e-\mu$ couplings of $H_1^0$ and $\Delta_3^+$. The $\tau-e(\mu)$ couplings of $\Delta L=0$ vector $H_{1}^0$ and $\Delta L=2$ scalar $\Delta_3$ with left-handed chirality are constrained by the absence of lepton-flavor-universality violation in $\tau$ decays, while the anomalous magnetic moment is most sensitive to an electroweak doublet scalar $H_2$ when only the neutral CP-even component is light.
The LEP measurement of $ee\to \mu\mu (\tau\tau)$ provides a complementary constraint on the $e-\mu (\tau)$ coupling. The $\Delta L=2$ vector boson $\Delta_2$ is currently best constrained by the anomalous magnetic moment of the muon.

Future experiments will improve the sensitivity to several of these observables. In particular, we expect that a future muonium-antimuonium conversion experiments will lead to a factor of $3$ improvement for the $y^{e\mu}$ coupling. Furthermore, the measurement of neutrino trident scattering at the DUNE near detector (and other neutrino detectors) will provide independent probes.

Despite of the expected success and the increase in sensitivity of low-energy precision experiments, the search for on-shell production of a bilepton at future lepton colliders will provide a complementary probe of CLFV couplings. Although low-energy precision constraints provide the strongest constraints for $\mu\tau$ final state for all bileptons apart from $\Delta_2$, future colliders can probe new parameter space for $e\ell$ ($\ell=\mu,\tau$) final states.
The FCC-ee with the highest integrated luminosity is the most sensitive machine and CEPC is the second most sensitive one in the low mass region. The CLIC and ILC with larger c.m.~energy can probe the high mass region for the new bileptons.

In summary, future lepton colliders provide complementary sensitivity to the CLFV couplings of bileptons compared with low-energy experiments. The future improvements of muonium-antimuonium conversion, lepton flavor universality in leptonic $\tau$ decays, electroweak precision observables and the anomalous magnetic moments of charged leptons will probe similar parameter space.

\acknowledgments
T.L. is supported by ``the Fundamental Research Funds for the Central Universities'', Nankai University (Grant Number 63191522, 63196013).


\bibliography{refs}
\end{document}